\documentclass[twocolumn,trackchanges]{aastex62} 

\usepackage{lineno}

\graphicspath{{./}{figures/}}

\received{???}
\revised{???}
\accepted{???}
\shorttitle{Mass and Age labels}
\shortauthors{Li \& Wang et al.}

\begin{document}

\title{Mass and Age determination of the LAMOST data with different Machine Learning methods}
\author{Qi-Da Li}
\affil{Department of Astronomy, China West Normal University, Nanchong, 637002, P.\,R.\,China}
\author[0000-0001-8459-1036]{Hai-Feng Wang}
\affil{Department of Astronomy, China West Normal University, Nanchong, 637002, P.\,R.\,China}
\affil{GEPI, Observatoire de Paris, Universit\'e PSL, CNRS, Place Jules Janssen 92195, Meudon, France}
\affil{CREF, Centro Ricerche Enrico Fermi, Via Panisperna 89A, I-00184, Roma, Italy}
\author{Yang-Ping Luo}
\affil{Department of Astronomy, China West Normal University, Nanchong, 637002, P.\,R.\,China}
\author{Qing Li}
\affil{Department of Astronomy, China West Normal University, Nanchong, 637002, P.\,R.\,China}
\author{Li-Cai Deng}
\affil{Department of Astronomy, China West Normal University, Nanchong, 637002, P.\,R.\,China}
\author{Yuan-Sen Ting}
\affil{Research School of Computer Science, Australian National University, Acton ACT 2601, Australia}

\correspondingauthor{HFW}
\email{hfwang@bao.ac.cn};\\

\begin{abstract}

We present a catalog of 948,216 stars with mass label and a catalog of 163,105 red clump (RC) stars with mass and age labels simultaneously. The training dataset is cross matched from the LAMOST (The Large Sky Area Multi-Object Fiber Spectroscopic Telescope) DR5 and high resolution asteroseismology data, mass and age are predicted by random forest method or convex hull algorithm. The stellar parameters with high correlation with mass and age are extracted and the test dataset shows that the median relative error of the prediction model for the mass of large sample is 3\% and meanwhile, the mass and age of red clump stars are 4\% and 7\%. We also compare the predicted age of red clump stars with the recent works and find that the final uncertainty of the RC sample could reach 18\% for age and 9\% for mass, in the meantime, final precision of the mass for large sample with different type of stars could reach 13\% without considering systematics, all these are implying that this method could be widely used in the future. Moreover, we explore the performance of different machine learning methods for our sample, including bayesian linear regression (BYS), gradient boosting decision Tree (GBDT), multilayer perceptron (MLP), multiple linear regression (MLR), random forest (RF) and support vector regression (SVR). Finally we find that the performance of nonlinear model is generally better than that of linear model, and the GBDT and RF methods are relatively better.

\end{abstract}

\keywords{Fundamental parameters of stars (555); Milky Way Galaxy (1054)}

\section{Introduction} 

To describe the current structure, evolution and formation history of the Milky Way, it is necessary to accurately estimate the mass and age of a large number of stars distributed throughout our home galaxy. Through the spectra of stars, astronomers could acquire many stellar parameters \citep{2017ApJS..229...30M,2019MNRAS.484.5315W,2020ApJS..249...29H,2020ApJS..246....9Z,2021arXiv210511624Z}. However, to date it is still not easy to acquire the age of stars accurately and precisely. The indirect isochrones method can obtain the age of clusters with relatively high precision by matching the observed data based on the stellar evolution model \citep{2010ARA&A..48..581S,2017ApJS..232....2X}, but for field stars the precision of this method might not be perfect due to that the high accurate stellar parameters must be needed. 

For a long time, due to the limitation of observation and data analysis, we can only estimate the ages of a small number of stars in the solar neighborhood \citep{1993A&A...275..101E,2004A&A...418..989N,2007ApJS..168..297T,2013A&A...560A.109H,2014A&A...565A..89B}. With the large sky surveys, such like LAMOST (The Large Sky Area Multi-Object Fiber Spectroscopic Telescope), \citet{2015RAA....15.1209X,2017ApJS..232....2X} estimated the ages of a large number of stars. Followed by this, it has been found that there is a relation between the carbon and nitrogen abundances and the ages of the giant stars, which has already been used to predict the ages of the red giant branch stars \citep{2016MNRAS.456.3655M,2016yCat..18230114N,2017ApJ...841...40H}. 

There are also other surveys that could provide the age of large sample. The GALAH survey (Galactic Archaeology with HERMES), a high-resolution spectroscopic survey, aims to the chemical tagging experiment \citep{2002ARA&A..40..487F,2010ApJ...713..166B}. For the very bright stars, more than 30 different elements can be measured and meanwhile, age, kinematic inventory of the solar neighbourhood has also been provided in \citet{2019A&A...624A..19B}. Bright giant stars become the primary targets for APOGEE survey (The  Apache Point Observatory Galactic Evolution Experiment), which is also a high-resolution spectroscopic survey with some works on the mass and age \citep{2013AJ....146...81Z,2016MNRAS.456.3655M,2017AJ....154...94M}. Recently, the precision of $\sim$5\% for mass and $\sim$20\% for age is acquired by \citet{2020ApJ...889L..34S} with TESS (Transiting Exoplanet Survey Satellite) data and meanwhile, there also are many other results with similar precision such as $\sim$6\% for mass and $\sim$20\% for age in \citet{2021arXiv210705831S} by using TESS asteroseismology of the Kepler red giants. And $\sim$10\% for mass and $\sim$30\% for age in \citet{2021MNRAS.502.1947M} with asteroseismology of giant stars in the TESS continuous viewing zones and beyond.

It has been found that there is a correlation between the age of solar-like stars and their surface rotation, and a detailed study has been carried out with asteroseismology data \citep{2014A&A...572A..34G,2014ApJS..211...24M,2016MNRAS.456..119C,2016Natur.529..181V}. At the present time, we all know asteroseismology is an effective method to estimate the mass and age of stars \citep{2011ApJ...730...63G,2014ApJS..210....1C}, however, it needs high-precision, long-time and high-resolution photometric observation so that, unfortunately, we still don't have large enough asteroseismological sample. 
 
Up to now, although there are many methods to predict the mass and age of stars, their precision and efficiency are still not perfect. We desperately need to make full use of big data to obtain more samples and try more methods to improve the prediction precision, thus then we can explore the assembly history of the Galaxy more effectively and more properties of the Milky Way mass distribution, population structure and dynamical evolution (e.g., \citep{2018MNRAS.477.2858W,2018MNRAS.478.3367W,2019ApJ...884..135W,2020MNRAS.491.2104W,2020ApJ...897..119W,2020ApJ...902...70W,2022arXiv220408542W,wang2022b,2019MNRAS.486.1167B,2021ApJ...922...80Y,Yang2022} and reference therein). 

Machine learning is a branch of artificial intelligence and we could make full use of high quality data for training through algorithm. By combining machine learning with high quality asteroseismology data, we could predict the relationship between stellar mass (age) and stellar parameters thus then we could get these two parameters of large sample with high confidence.

During this paper, we use novel machine learning method to estimate mass of a larger sample and a smaller sample for age and mass of red clump stars in LAMOST. Furthermore, we compare the different machine learning methods quantitatively for the first time. 

The paper is structured as follows: Section 2 presents the data we adopt, Section 3 is the method introduction we use, Section 4 shows our results, Section 5 is for discussion, and finally Section 6 gives a brief summary of our work.

\section{Data} 
\subsection{Catalogs}
\citet{2019ApJS..245...34X} has provided 8,162,566 stars from LAMOST survey and the chemical abundances are derived from DD-Payne model, which is inherited from both the Payne \citep{2019ApJ...879...69T} and the Cannon \citep{2015ApJ...808...16N}. In this work, we use this catalog to obtain the chemical abundances of stars.

 \citet{2018ApJ...858L...7T} has provided us 175,202 red clump stars in LAMOST with 3\% contamination, and also includes two asteroseismology parameters $\Delta$P and $\Delta\nu$. We use this catalog to obtain RC stellar label and notice the $\Delta$P and $\Delta\nu$ are also obtained from stellar spectra, the frequency separation ($\Delta\nu$) between adjacent acoustic p-modes and the period spacing ($\Delta$P) of the mixed gravity g- and acoustic p-modes could be used for the separation of red clump stars and red giant branch stars\citep{2018ApJ...858L...7T,2018ApJ...853...20H}. The precision for LAMOST $\Delta$P and $\Delta\nu$ is 50\,s and 1$\mu$HZ respectively, which is enough for the age/mass determination according to the previous results \citep{2018ApJ...858L...7T,2019ApJ...879...69T}. During this work, we determine the final age and mass using new training dataset and new methods we choose, then we compare with other catalogs in order to test the robustness of the different methods.

\citet{2018ApJS..239...32P} has provided ages of 6,676 stars in APOKASC-2, which are derived from the model of \citet{Serenelli et al.(2018)} using mass, radius, [Fe/H] and [$\alpha$/Fe]. We train our model for mass and age by this high quality high resolution asteroseismology catalog. To be more specific, this catalog is stellar properties for a large sample of evolved stars with APOGEE spectroscopic parameters and Kepler asteroseismic with the help of five independent techniques, the median random mass uncertainties for red giant branch (RGB) stars could reach 4\%, for RC stars could reach 9\% level and meanwhile the age precision is within 8\% respectively, which is suitable for training sample.

In short, thanks to these works above we use chemical abundance from \citet{2019ApJS..245...34X}, precise mass and age from \citet{2018ApJS..239...32P}, red clump label and $\Delta$P and $\Delta\nu$ from \citet{2019ApJ...879...69T}. Then we use the new machine learning methods and new high quality asteroseismic age and mass, to estimate the mass of large sample for \citet{2019ApJS..245...34X}, and red clump stars age \& mass for \citet{2019ApJ...879...69T}.

\begin{figure*}
  \centering
  \includegraphics[width=0.9\textwidth]{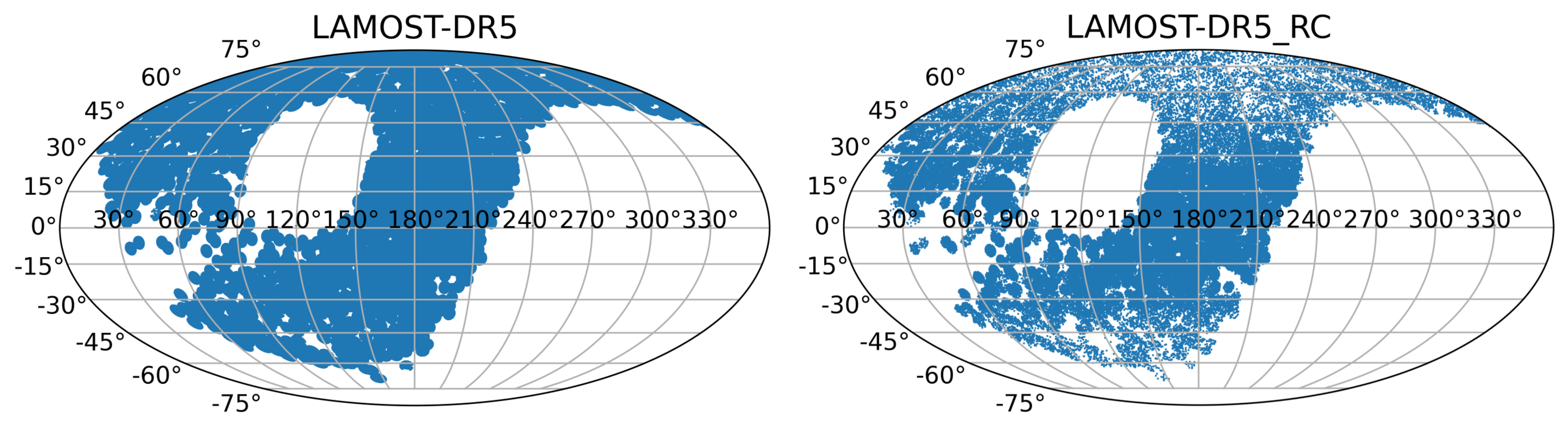}
  \caption{Distribution of the LAMOST data for mass estimation we use (left panel) and the right one is the RC distribution we use for age \& mass.}
  \label{RA - DEC_LAMOST}
\end{figure*}

After cross matching the above catalogs, we firstly get 4,479 stars to predict large sample mass (LS-mass), and 1,806 stars for red clump mass (RC-mass) and red clump age (RC-age), notice that these are not the final dataset as shown in the next part. The distribution of sample needed to be predicted in the galactic longitude and latitude of the celestial coordinates is shown in Fig.~\ref{RA - DEC_LAMOST}.

\subsection{Final training datasets}

In order to improve the precision of machine learning prediction, we do the following experiment for the three catalogs mentioned above. 

\begin{figure*}
  \centering
  \includegraphics[width=0.9\textwidth]{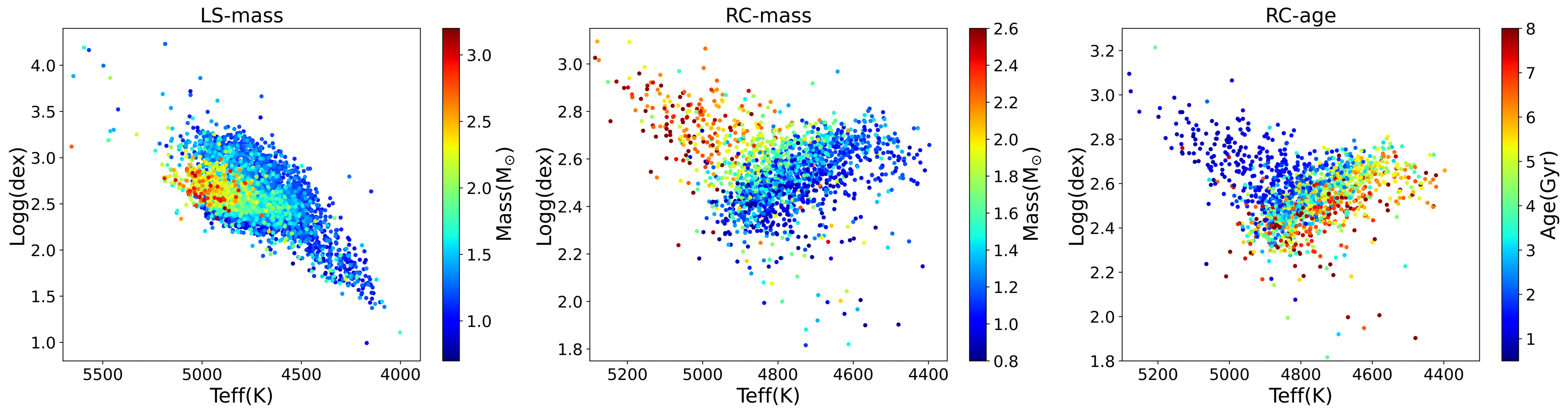}
  \caption{The final training sample distribution for mass and age on the Teff and log g plane.}
  \label{Training-Teff-Logg-mass-age}
\end{figure*}

The three datasets after the first cross match mentioned above are separated as the test and training sample equally, then we firstly use RF to train and make mass and age prediction for the test dataset. For large sample stars (LS-mass), we select stars whose absolute error of mass prediction is less than 1 M$_{\sun}$ and relative error is less than 0.3, and for RC stars, we select stars whose absolute error of mass (age) prediction is less than 1 M$_{\sun}$ (3 Gyr) and relative error is less than 0.4. Notice that here we only use 200 decision trees and make full use of all stellar parameters shown in Fig.~\ref{Feature_importances} as input in the method to finish this step. After this, we finally get LS-mass set 4,246 stars, RC-mass set 1,751 stars and RC-age set 1,384 stars for training and predicting, as detailed in Fig.~\ref{Training-Teff-Logg-mass-age}, which is showing the final training mass and age distribution on the Teff-log g plane.

\section{Method} 

\subsection{Feature importance}

The machine learning methods used in this paper are mainly from Scikit-learn (sklearn) \citep{2019arXiv191006853A,2019arXiv191101217M,2020arXiv200511251F}, which can be divided into six categories: classification, regression, clustering, dimensionality reduction, model selection, preprocessing.

\begin{figure}
  \includegraphics[width=0.4\textwidth]{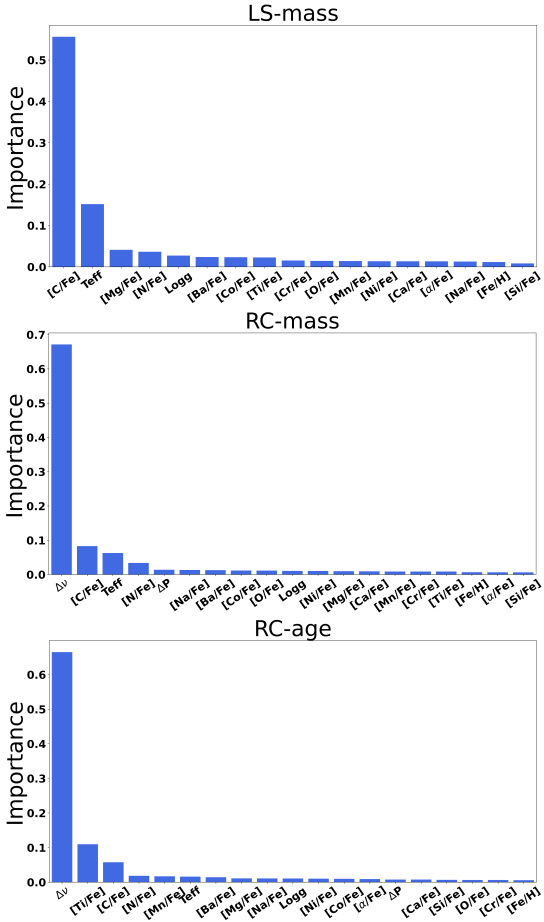}
  \caption{The results of feature extraction using random forest. The different figures are three different training samples that we have selected, the top one is for the large sample containing different type of stars, for which we only estimate mass, the middle and bottom one are for red clump stars, for which we could estimate mass and age. The importance represents the contribution of the stellar parameter to our prediction model, it is actually the relative importance.}
  \label{Feature_importances}
\end{figure}

Firstly, we explore the feature importance distribution of the stellar parameters for the mass/age of the three selected training samples with RF method shown in Fig.~\ref{Feature_importances}. 
In order to avoid the severe impact of one feature on the prediction due to the dimension problems unexpected, we choose to do the standardization for different features which can accelerate the convergence of weight parameters. Standardization or Z-score normalization is the transformation of features by subtracting from mean and dividing by standard deviation.

The RF method adopted here is based on decision trees and the final prediction result is also dependent on these trees. The correlation between different parameters can be easily identified with the help of information gain used to train the model so this method have good robustness and overfitting could be avoided.

The importance is implying the relative significance and we have a test to find that the importance of many stellar parameters are highly correlated so it is reasonable that we choose to use the first six or nine parameters to estimate mass and age. As shown in Equation 3 in \citet{2018ApJS..239...32P}, the mass is very sensitive to the $\Delta\nu$ and we all know the age is also sensitive to mass, so it is not strange to see the $\Delta\nu$ is the most important factor for the RC age and mass.

\subsection{Features choice}

\begin{figure*}
  \centering
  \includegraphics[width=0.8\textwidth]{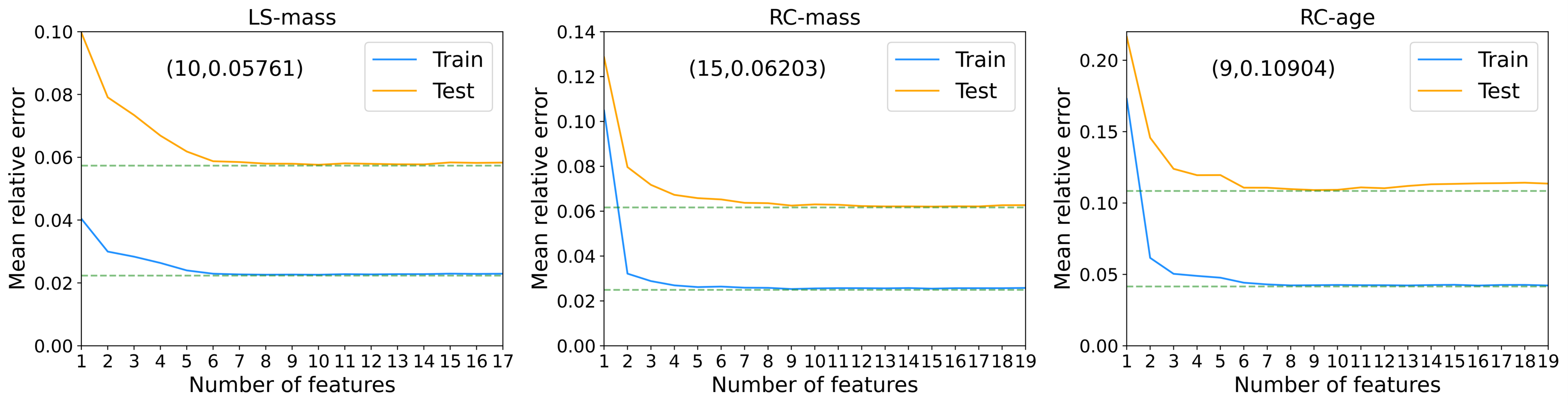}
  \caption{The relations between the number of training features and the mean relative error of the test dataset in our prediction model. Different figures are different samples. The blue line represents the training dataset, the orange line represents the test dataset, and the green dotted lines represents the minimum value used to guide our eyes for the stable pattern. The minimum values are also labelled on the top of each figure. Notice that it is the minimum value but not the final feature we adopt, we choose final features empirically and accordingly.}
  \label{RF_NOF - MRE}
\end{figure*}

The relation between the precision of prediction and the number of features in the training dataset, base on the relative errors distribution vs. feature numbers, is clearly shown in Fig.~\ref{RF_NOF - MRE}. The mean relative error of the test dataset decreases with the increase of the number of training features (orange line) until stable pattern. Based on this pattern, we choose the first six stellar parameters to train the model for mass of large sample stars (LS-mass) and meanwhile first nine features for mass and six features for age of RC stars respectively.

\begin{figure}
  \centering
  \includegraphics[width=0.45\textwidth]{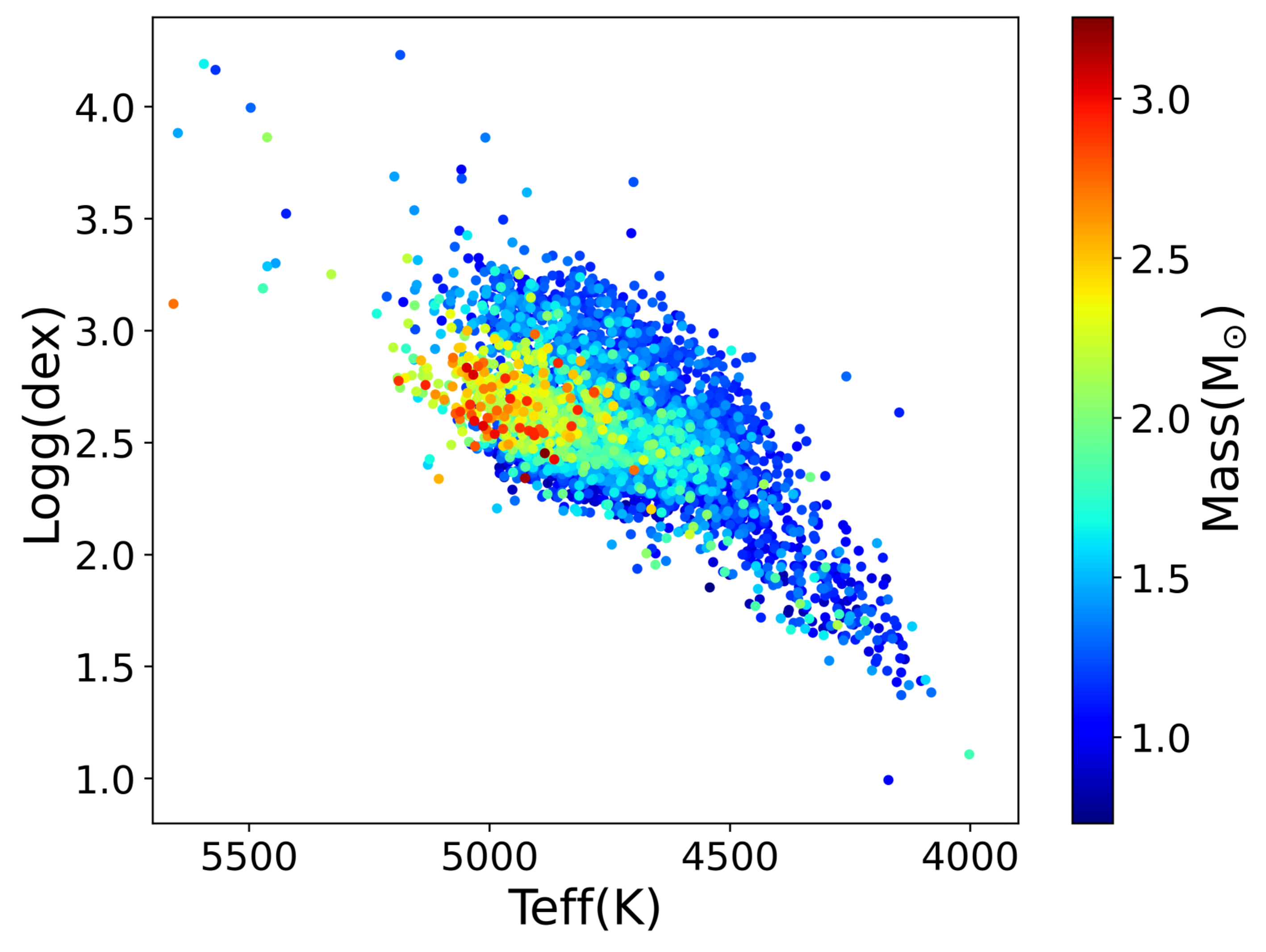}
  \caption{Mass distribution of the stars we use to create convex hull and train prediction models, on the Teff and logg plane, different colours represent different mass.}
  \label{Teff - Logg - Mass}
\end{figure}

We notice that the LS-mass are mixed with different types of stars which might not belong to the training dataset, so we use the first six stellar parameters of [C/Fe], Teff, [Mg/Fe], [N/Fe], log g, [Ba/Fe] to construct a convex hull in order to determine which stellar types our training model are suitable for, as displayed in Fig.~\ref{Teff - Logg - Mass} , we could see our sample is mainly consist of K giant stars including RC and RGB, and there are also very few possible other type stars as such G type, which is consistent with the result that APOKASC is mainly consist of RGB and RC. So our large sample is almost consist of K giants, we find that LAMOST DR5 contains around 1 million K giant stars, in this work we also use convex hulls to select 948,216 stars, which are self-consistent. Notice that our method could be wildly used in different type of stars if the quality and quantity of the training dataset is enough in the future, and in order to avoid the mixing effects of RGB and RC, we choose not to estimate the age of all large sample here, age of the RGB estimation will be shown in the next work. Algorithms that construct convex hulls of various objects have been wildly used in astrophysics, mathematics and computer science.  Finally, we have 948,216 stars for LS-Mass suitable for the training model based on the \citet{2018ApJS..239...32P}. Notice that we have also removed some vacancy values for the red clump catalog before mass and age determination, then we finally get the 163,105 stars to be predicted without convex hull algorithm.

\section{Results} 

\subsection{Final age and mass distribution} 

\begin{figure}
  \centering
  \includegraphics[width=0.45\textwidth]{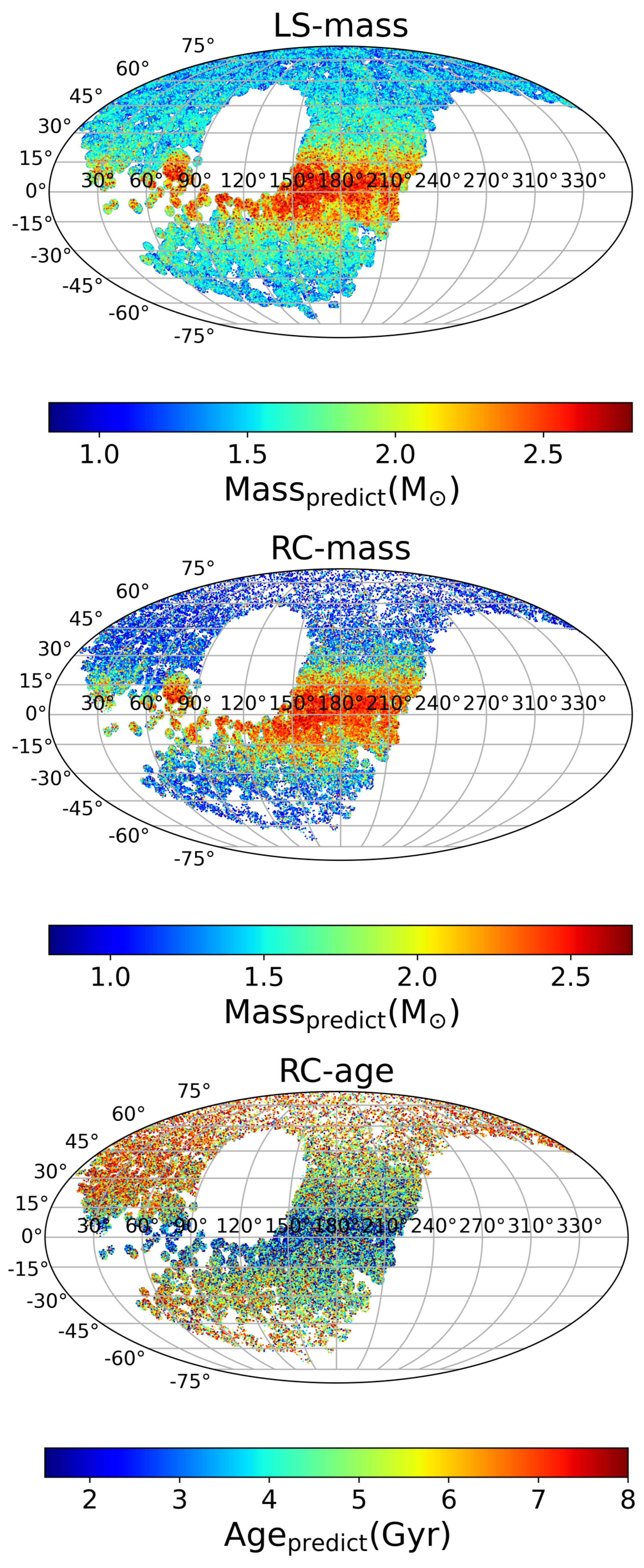}
  \caption{The distribution of predicted mass (age) in celestial sphere coordinates. The top one is the mass distribution of large sample, the middle one is the mass distribution of red clump stars, the bottom one is the age distribution of red clump stars.}
  \label{RA - DEC}
\end{figure}

The final predicting mass of 948,216 stars (using [C/Fe], Teff, [Mg/Fe], [N/Fe], log g, [Ba/Fe] ), mass (using $\Delta\nu$, [C/Fe], Teff, [N/Fe], $\Delta P$, [Na/Fe], [Ba/Fe], [Co/Fe], [O/Fe]) and age (using $\Delta\nu$, [Ti/Fe], [C/Fe], [N/Fe], [Mn/Fe], Teff) of 163,105 RC stars are presented vividly in Fig.~\ref{RA - DEC}, coloured by the mass or age on the Galactic longitude and latitude celestial sphere. For the mass distribution, we could see the more massive stars are located in the disk similar to the mass pattern for the red clump stars in the middle panel, and the age distribution of red clump stars is also showing the younger stars are mainly located in the low latitude. It could be naturally understood that there are more star forming regions in the disk so that more massive stars and younger stars located in the disk and low latitude.  

\begin{figure*}
  \centering
  \includegraphics[width=0.95\textwidth]{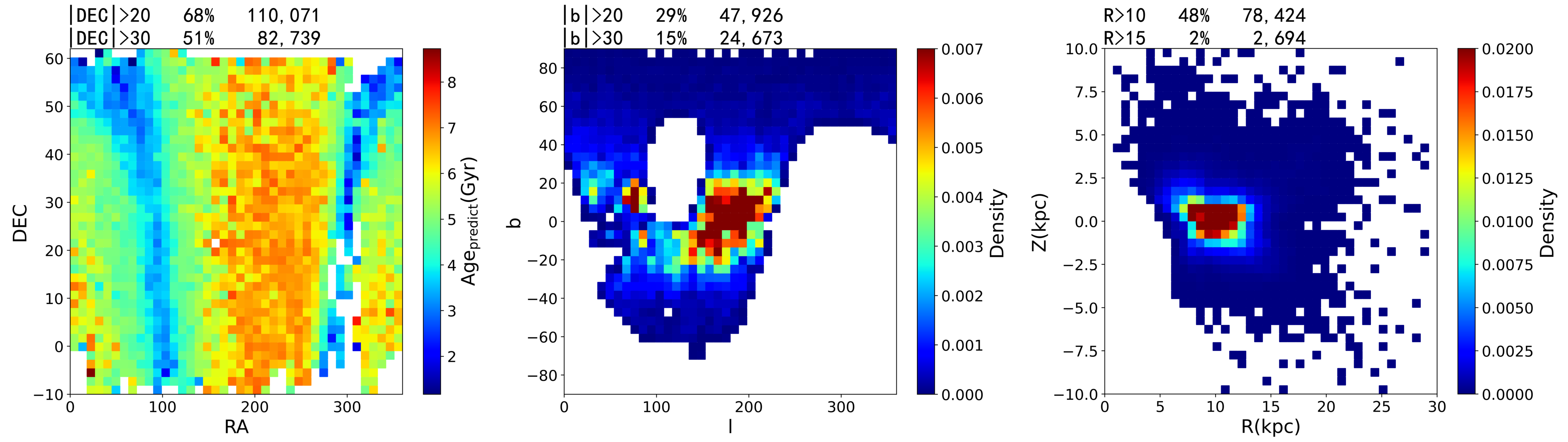}
  \caption{The distribution of RC-age on the RA vs. DEC plane is shown in the left panel coloured by age, the number and fraction for declination beyond 20 or 30 degree are denoted on the top of the panel. The middle one is for density distribution in the longitude and latitude plane, star counts and fraction beyond 20 or 30 degree for latitude are labeled on the top of this figure. The right panel is R and Z plane in cylindrical Galactic coordinates coloured by density/star counts and fraction larger than 10 or 15 \,kpc for radial distance are also denoted on the top.}
  \label{Density}
\end{figure*}

The distribution of age in the right ascension and declination plane is also shown in the left panel of Fig.~\ref{Density}, the number and fraction for declination beyond 20 or 30 degree are denoted on the top, they are 110071 and 68\%, 82739 and 51\% respectively. The middle one of this figure is for density distribution in the longitude and latitude plane, star counts and fraction beyond 20 or 30 degree for latitude are labeled on the top of this figure, they are 47,926 and 29\%, 24,673 and 15\% respectively. The right panel in this figure is R and Z plane in cylindrical Galactic coordinates coloured by density/stellar number and fraction larger than 10 or 15 \,kpc for distance are also denoted on the top, they are 78,424 and 48\%, 2,694 and  2\% separately. 

\begin{figure*}
  \centering
  \includegraphics[width=0.6\textwidth]{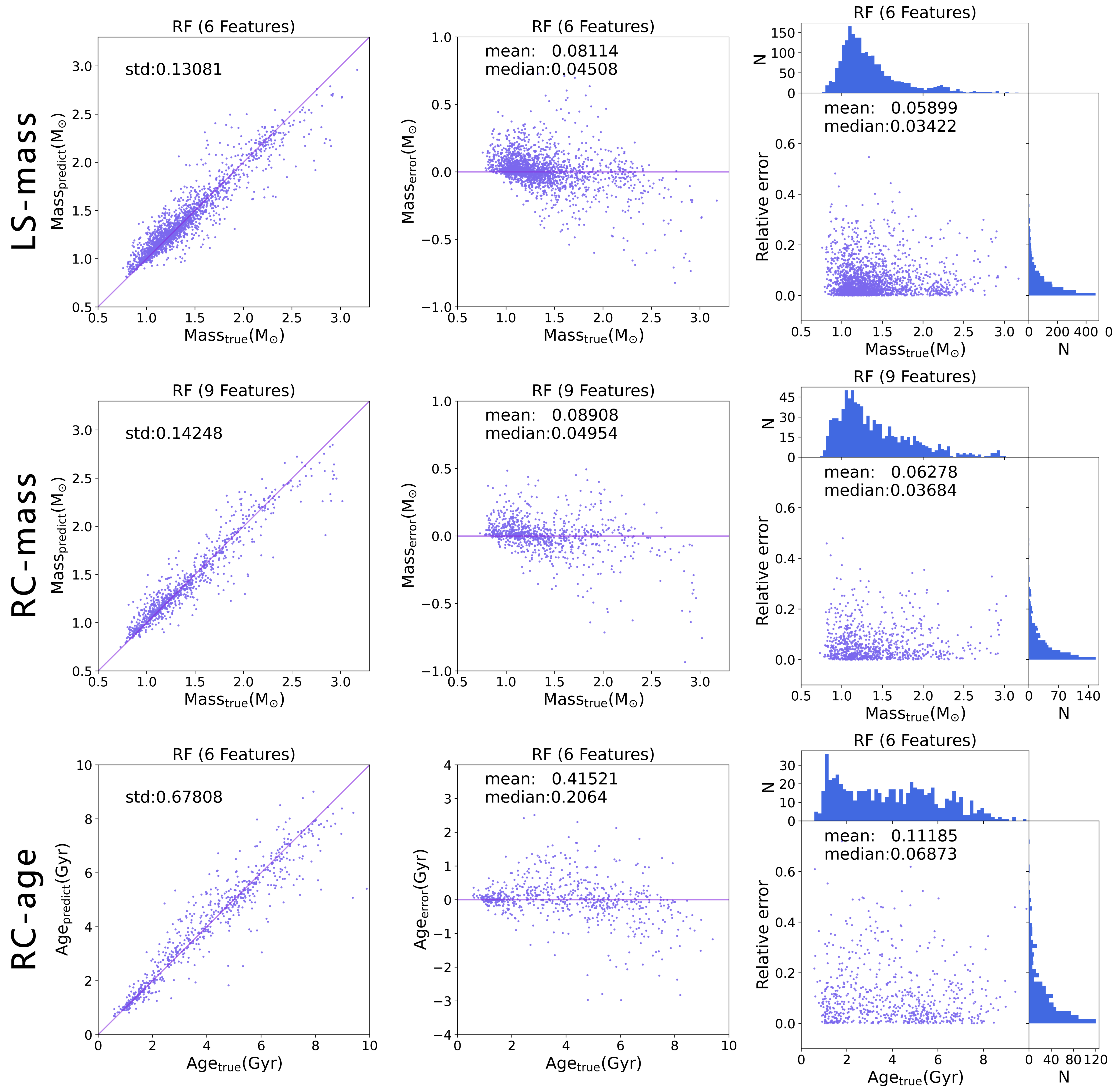}
  \caption{The predicted results of our test datasets using RF. Different rows represent different groups of samples, and different columns show the dispersion (M$_{\sun}$, Gyr), absolute error (M$_{\sun}$, Gyr) and relative error respectively. Dispersion means the standard deviation of the predicted ages/mass minus the true values in the catalog we used, the absolute error is the predicted value minus the true value, the relative error is the predicted value minus the true value divided by true value, notice in this work we use the relative error for the final precision uniformly. The dispersion, mean and median values of the data are marked in the upper left corner of each figure, and the final number of features we adopt to train each model is marked at the top of each panel.}
  \label{RF_prediction}
\end{figure*}

Fig.~\ref{RF_prediction} shows the results of our method for the test datasets of three groups, from the top left to the top right, the y-axis is predicted mass, the absolute mass error and the relative error, the x-axis is the true mass from asteroseismology.  As shown in the figure, the predicted dispersion of the large sample mass is 0.13 M$_{\sun}$, the mean absolute error is 0.08 M$_{\sun}$ and the median is 0.05 M$_{\sun}$; the mean relative error is 6\% and the median is 3\%. Dispersion means the standard deviation of the predicted age/mass minus the true values in the catalog we used, the absolute error is the predicted value minus the true value, and the relative error is the predicted value minus the true value divided by true value, notice in this work we use the median relative error for the final precision uniformly.

Similarly, the middle of Fig.~\ref{RF_prediction} is the red clump stars mass, as shown in the label, the predicted dispersion of mass of RC stars is 0.14 M$_{\sun}$, the mean absolute error is 0.09 M$_{\sun}$ and the median value is 0.05 M$_{\sun}$, the mean relative error is 6\% and the median value is 4\%.  

It can be found in Fig.~\ref{RF_prediction} for the prediction of mass, the precision of RC stars (4\%) is slightly worse than that of large sample stars (3\%) for the test dataset. The main reason is that the number of stars in the training samples is different. The larger the sample size, the more effectively the machine learning method could find the rule. Moreover, the  predicted dispersion of age of RC stars is 0.68 \,Gyr, the mean absolute error is 0.42 \,Gyr and the median value is 0.21 \,Gyr, the mean relative error is 11\% and the median relative value is 7\%. We could speculate that the precision of age of RC could reach higher if we have higher quality catalog.

\begin{figure}
  \includegraphics[width=0.4\textwidth]{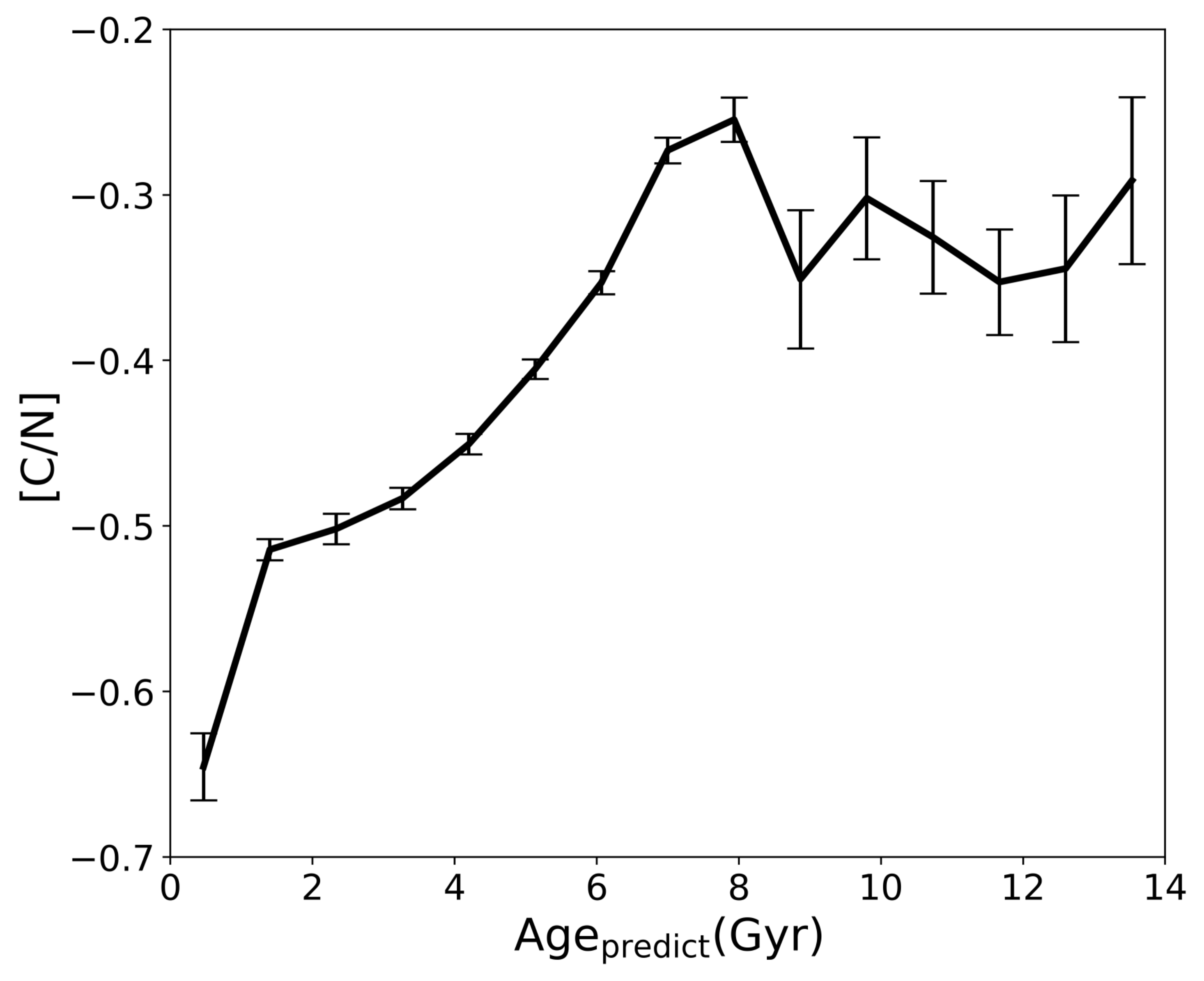}
  \caption{The relationship between the predicted age and [C/N], the black line is the median value in each bin with poission error.}
  \label{Age - C_N - Counts}
\end{figure}

Then we explore the relations between the predicted age and [C/N], as shown in Fig.~\ref{Age - C_N - Counts}. We could see that in the region where age is less than or equal to 8 \,Gyr, the age and [C/N] show a good linear relationship, which is consistent with our expectation. While in the region where age is older than 8 \,Gyr, it seems that there is no obvious pattern due to that the RC stars are inclined to the relatively younger group, and the number of old stars is very small in our sample, so it is impossible to make high-precision statistics.

\subsection{More comparisons} 

\begin{figure*}
  \centering
  \includegraphics[width=0.9\textwidth]{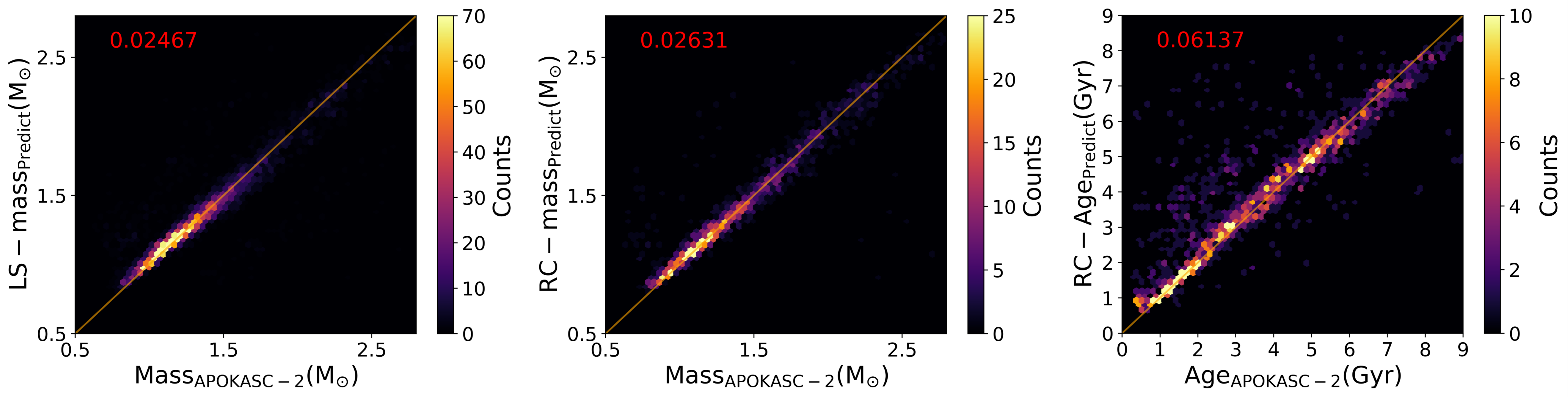}
  \caption{The comparison between the mass and age we predict and the reference values we use during this work, and the number marked on the figure represents the median value of relative error for our method. It is consist of the common stars of LAMOST data we predict and APOKASC-2 in this work. The purpose here is method validation and the precision is naturally quite good for this dataset since we use the APOKASC-2 for training.}
  \label{Prediction - APOKASC-2}
\end{figure*}

\begin{figure*}
  \centering
  \includegraphics[width=0.9\textwidth]{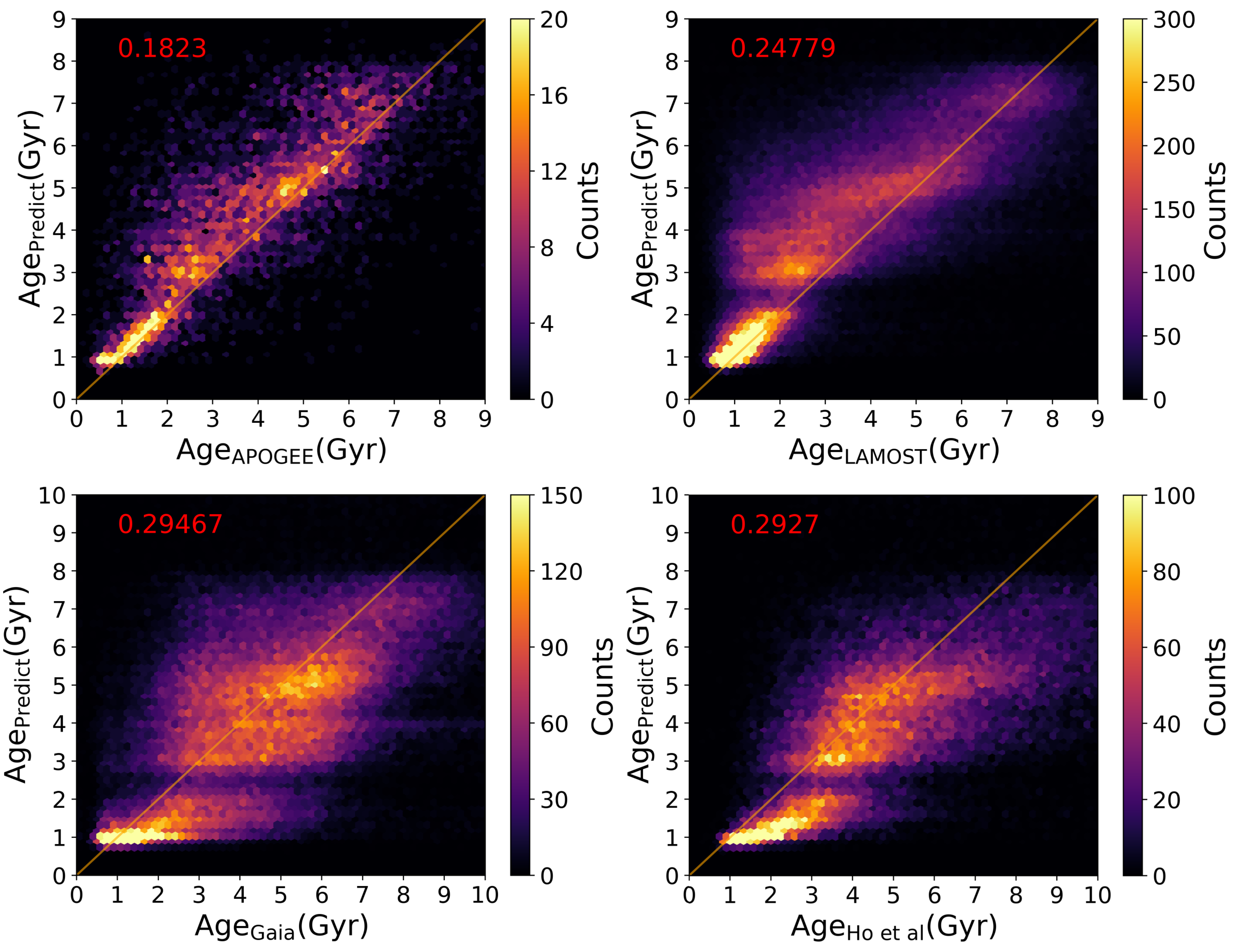}
  \caption{Comparing our predicted age with other works using LAMOST, APOGEE and Gaia data. On the top left is the age of APOGEE data using different method \citep{2019ApJ...878...21T}, the top right one is the age of LAMOST data\citep{2018ApJ...858L...7T}, the bottom left is the age of Gaia data \citep{2018MNRAS.481.4093S}, and the last one is the work of \citet{2017ApJ...841...40H}. The median value of relative error is shown on the top left and the consistency is acceptable. We have fewer stars around 2 Gyr in the training dataset so there are apparently disconnect features.}
  \label{Age_comparison}
\end{figure*}

\begin{figure*}
  \centering
  \includegraphics[width=0.85\textwidth]{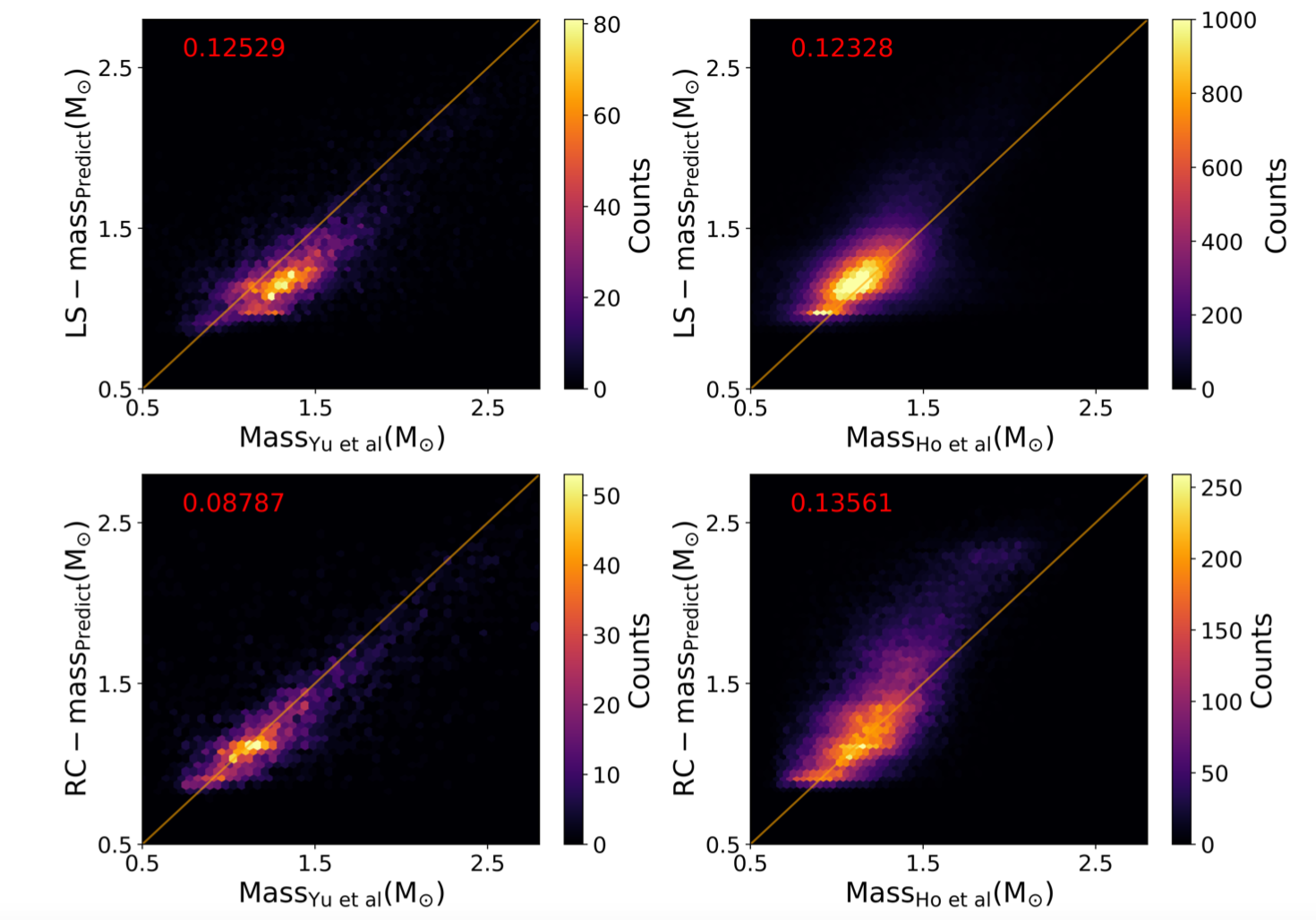}
  \caption{The figure is showing the mass comparisons between this work and others of \citet{2018APJS....236...42} and \citet{2017ApJ...841...40H}.The median value of relative error is shown on the top left and the consistency is acceptable.}
  \label{Mass_comparison}
\end{figure*}

Fig.~\ref{Prediction - APOKASC-2} shows the comparison between the mass or age we predict and the reference values we use, the consistency provides verification for the robustness of our method. We also compare our predicted age with other works based on LAMOST, APOGEE and Gaia data, which will also provide independent verification for the method. The comparison results are shown in Fig.~\ref{Age_comparison} and we could see, the top left one is comparison for the common stars of APOGEE \citep{2019ApJ...878...21T}, the top right one is for LAMOST \citep{2018ApJ...858L...7T} \footnote {Age is not shown clearly in the paper but it is determined simultaneously.}, the bottom left one is for Gaia \citep{2018MNRAS.481.4093S}, and the last one is for the work of \citet{2017ApJ...841...40H}. We could see although there are some differences, for the overall trend the consistency is acceptable. Similarly, the first four subfigures of Fig.~\ref{Mass_comparison} show the mass comparisons for other works. The left panel is compared to \citet{2018APJS....236...42}, right panel is compared to \citet{2017ApJ...841...40H}, top panel is for LS-mass, and the bottom panel is for RC-mass, all are matched well with some reasonable difference. 

Compared with APOGEE high quality data we could claim for this work, the precision of RC age could reach 18\% (top left in Fig.~\ref{Age_comparison}) and by matching with the high precision Kelpler asteroseismology data we could claim our uncertainty of RC mass could reach 9\% (bottom left of Fig.~\ref{Mass_comparison}). Meanwhile, the precision of LS-mass could be 13\% (top left of Fig.~\ref{Mass_comparison}). All these final precision are based on the final relative error analysis using high precision asteroseismology dataset and we frankly admit that the systematics might be ignored so we need more works in the future.

\begin{figure*}
  \centering
  \includegraphics[width=0.5\textwidth]{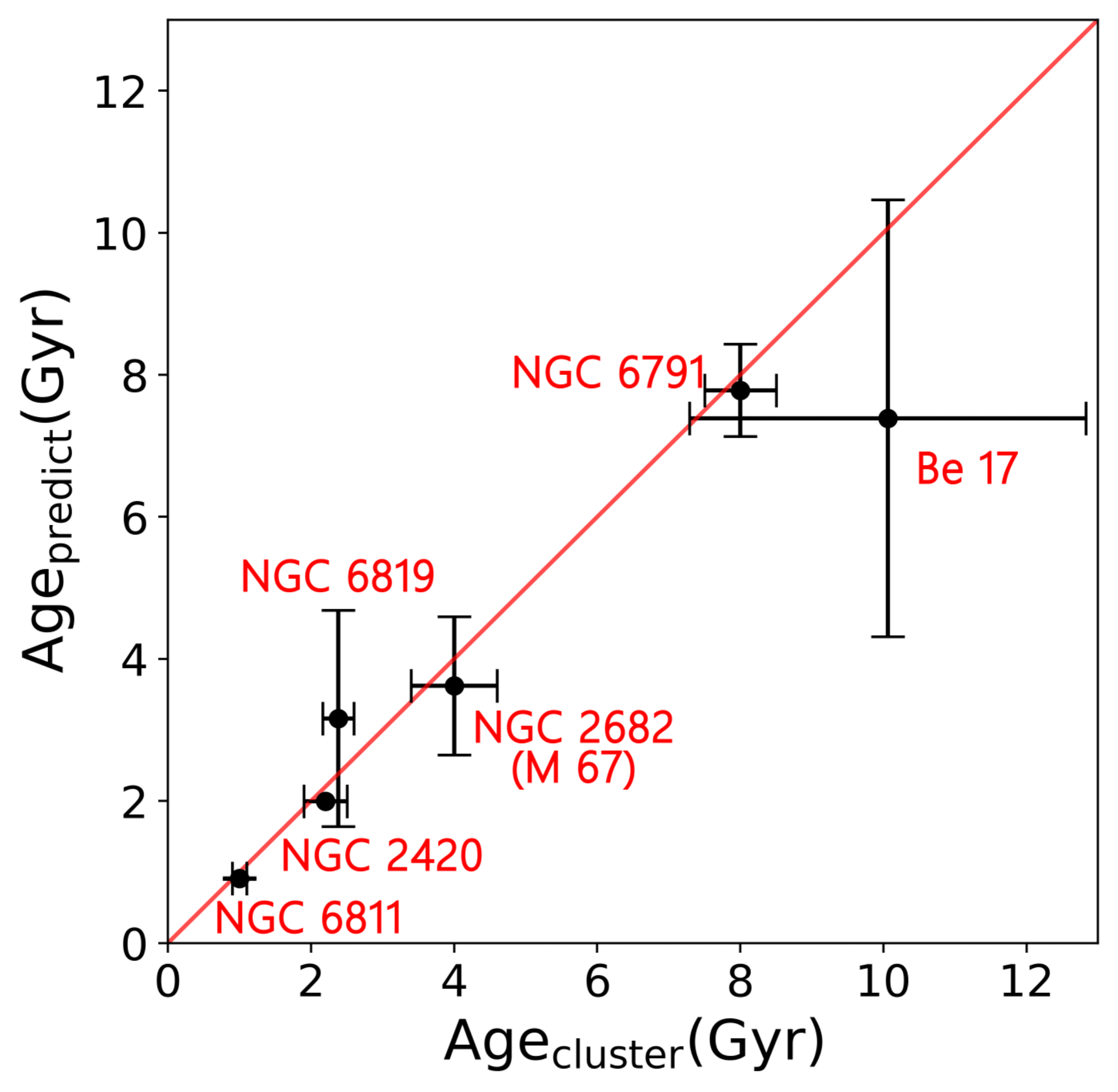}
  \caption{Open clusters comparisons for our determinations and literature values based on \citet{2013AJ....145....7J,2011AJ....142...59J,2016AJ....151...66B,2016ApJ...832..133S,2008AJ....135.2264G,2008A&A...492..171G,2006MNRAS.368.1971B} and reference therein. The comparison is quite well except the older last cluster and the error bar is calculated by the Gaussian dispersion or literature values.}
  \label{cluster_comparison}
\end{figure*}

Moreover, we also compare the open cluster (OC) age using our final sample, the OC is chosen by the spatial locations, kinematics (line of sight velocity, proper motions)and metallicity clustering distributions. As we could see in 
Fig.~\ref{cluster_comparison} the relative errors are NGC 6811: 9.1\%, NGC 2420: 9.3\%, NGC 6819: 23.4\%, NGC 2682: 9.5\%, NGC 6791: 2.7\%, Be 17: 33.5\%. The final median relative error is 9.5\%, which strongly supports our final conclusions. Notice that we use our final LAMOST RC catalog to select OC memberships and then compare with literature values. In our final RC catalog, the stellar number of memberships for these open clusters mentioned above is: NGC 6811: 2, NGC 2420: 1, NGC 6819: 6, NGC 2682: 2, NGC 6791: 2, Be 17: 4.

\begin{figure*}
  \centering
  \includegraphics[width=0.8\textwidth]{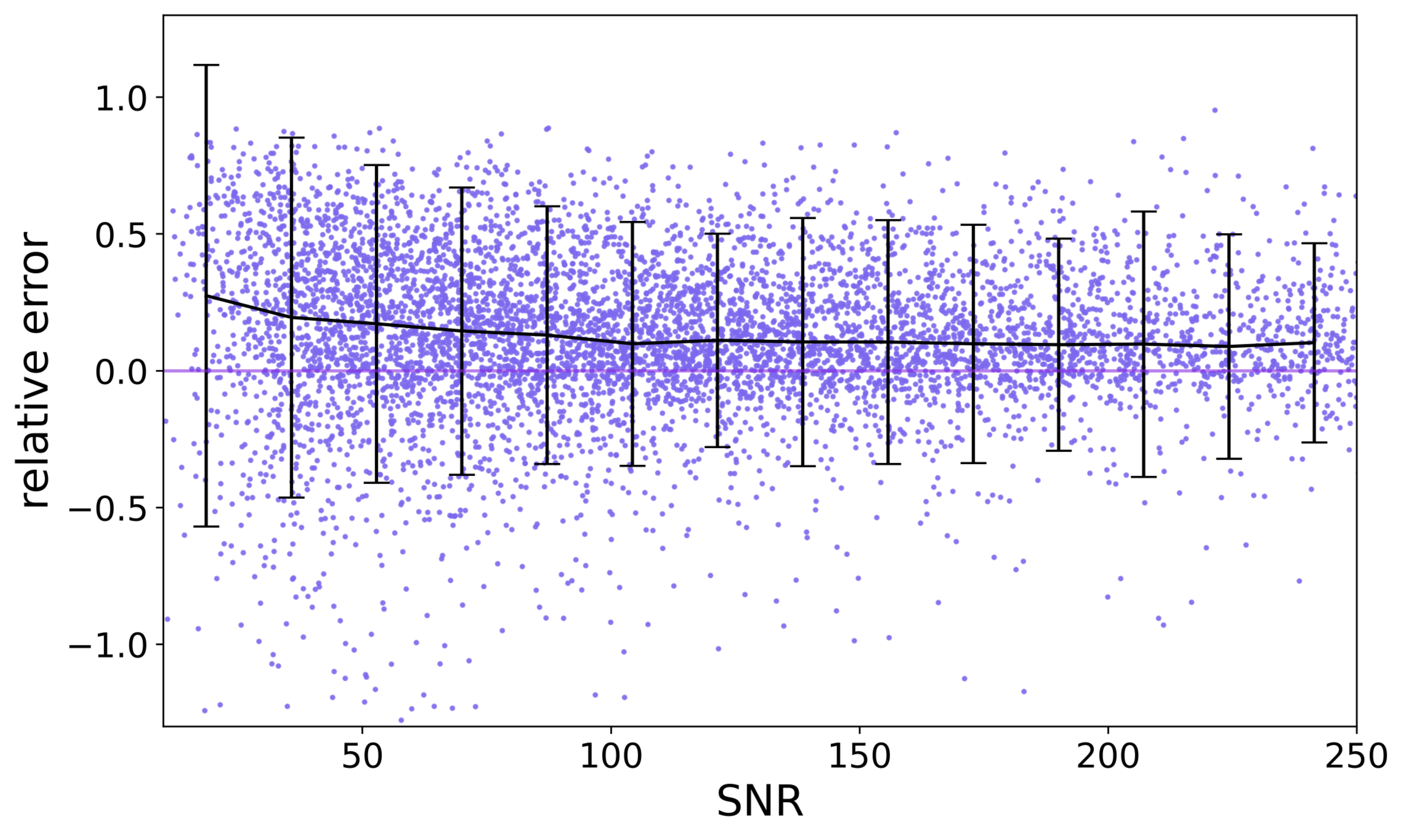}
  \caption{The relative error of RC age vs. SNR during this work, and the error bar represents the standard deviation in each bin.}
  \label{SNR - RE}
\end{figure*}

We also explore the relationship between RC-age relative error and SNR (the ratio of the intensity of a signal to the background noise detected by a measuring instrument for spectra used for LAMOST stellar parameters estimation), as shown in Fig.~\ref{SNR - RE}, the relative error tends to be stable with the increase of SNR.The distributions of the relative errors of mass and age for our test dataset with stellar parameters Teff, log g and [Fe/H] are also displayed in Fig.~\ref{Teff,Logg,FeH - Mass,Age}, which is showing that the robustness of our method with small dispersion.

\begin{figure*}
  \centering
  \includegraphics[width=0.7\textwidth]{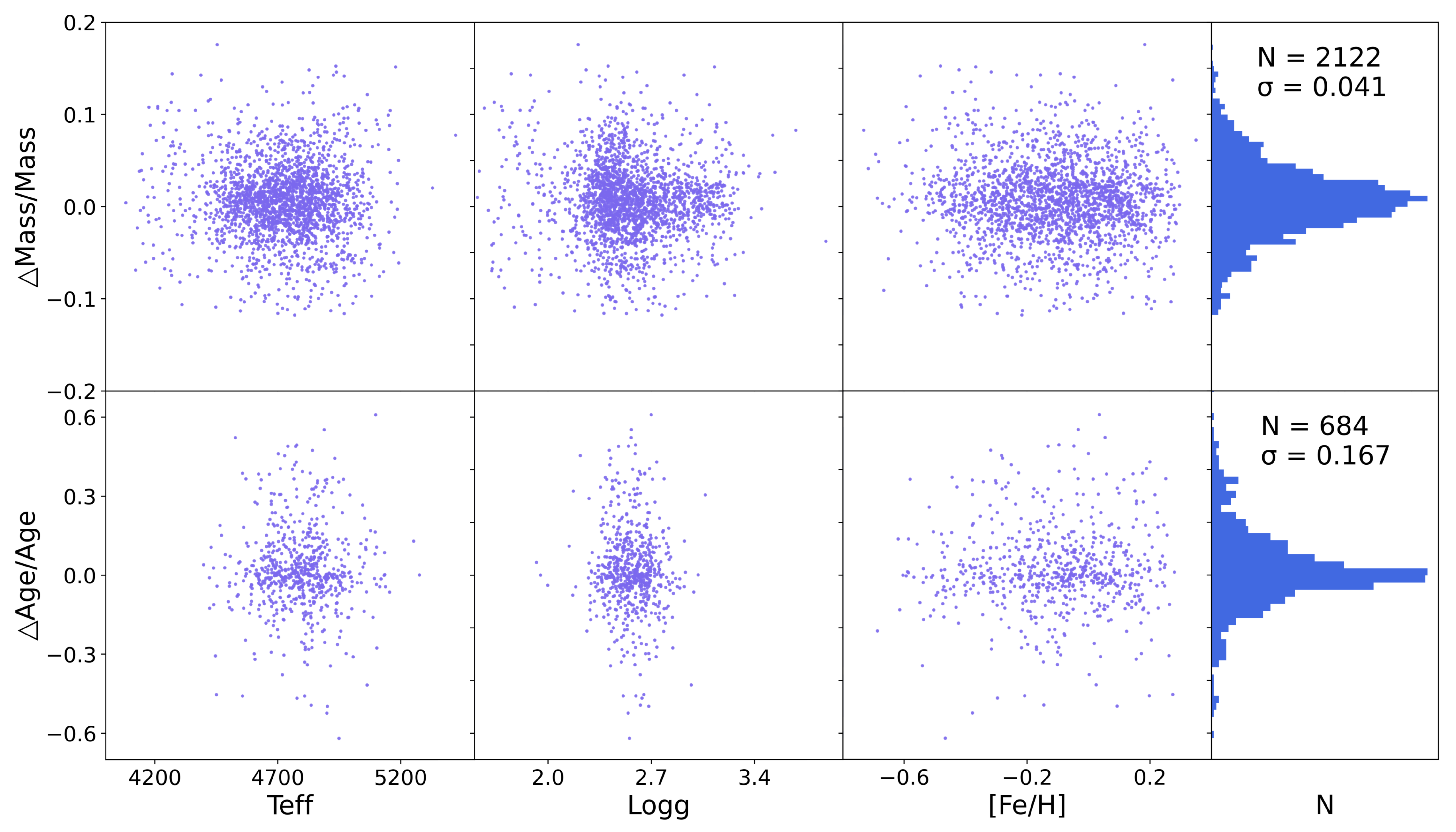}
  \caption{The distribution of relative errors of LS-mass and RC-age for our test dataset we predict along with Teff (K), logg (dex) and [Fe/H] (dex).}
  \label{Teff,Logg,FeH - Mass,Age}
\end{figure*}

\begin{figure*}
  \centering
  \includegraphics[width=0.8\textwidth]{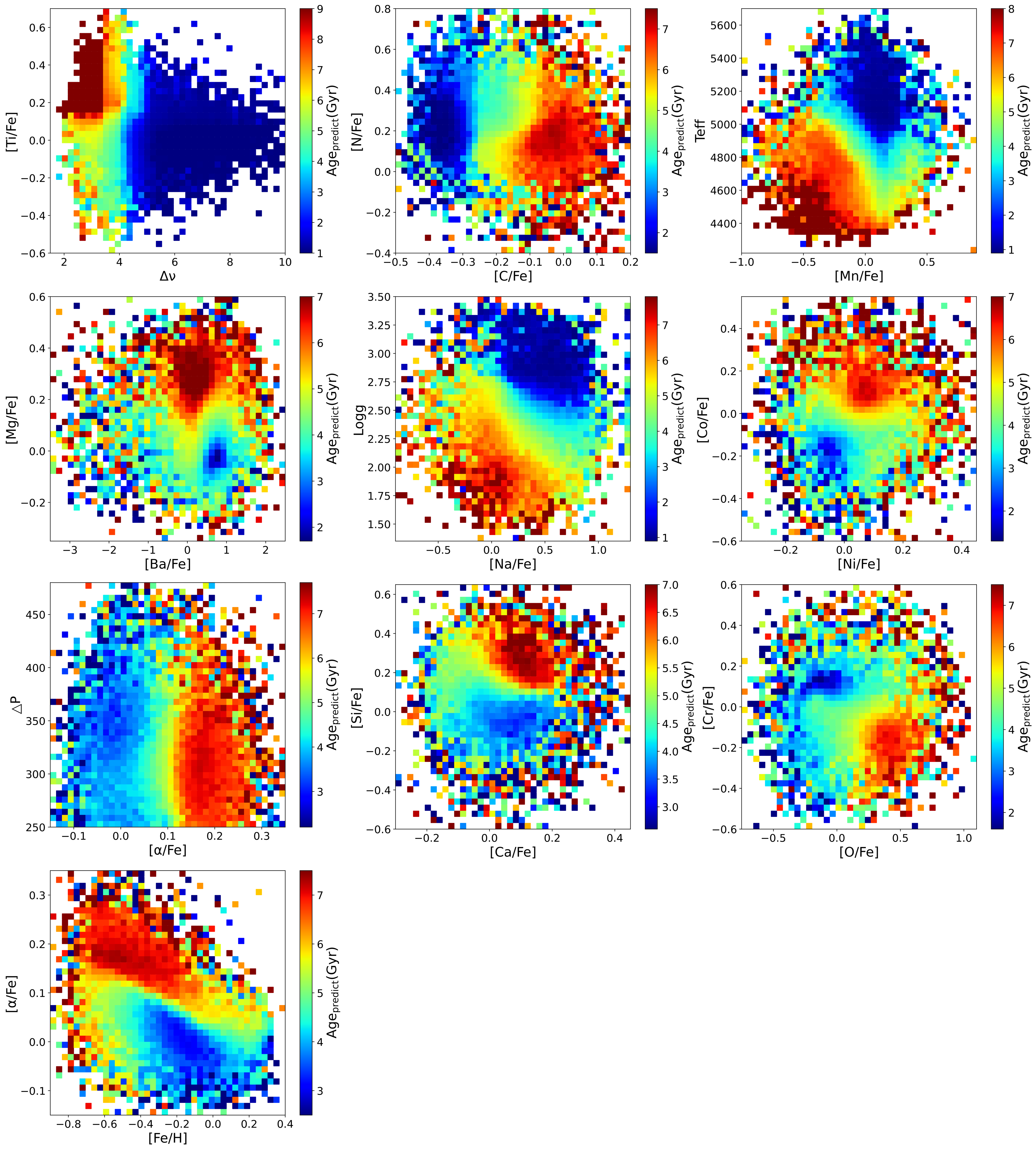}
  \caption{The distribution of our predicted ages over every two stellar parameters. From the top left to the bottom left are: $\Delta\nu$ vs. [Ti/Fe], [C/Fe] vs. [N/Fe], [Mn/Fe] vs. Teff, [Ba/Fe] vs. [Mg/Fe], [Na/Fe] vs. $logg$, [Ni/Fe] vs. [Co/Fe], [$\alpha$/Fe] vs. $\Delta$P, [Ca/Fe] vs. [Si/Fe], [O/Fe] vs. [Cr/Fe], [Fe/H] vs. [$\alpha$/Fe].}
  \label{Feature_parameter}
\end{figure*}

Fig.~\ref{Feature_parameter} shows the age distribution on the panel of different stellar parameters, From the top left to the bottom left are: $\Delta\nu$ vs. [Ti/Fe], [C/Fe] vs. [N/Fe], [Mn/Fe] vs. Teff, [Ba/Fe] vs. [Mg/Fe], [Na/Fe] vs. log g, [Ni/Fe] vs. [Co/Fe], [$\alpha$/Fe] vs. $\Delta$P, [Ca/Fe] vs. [Si/Fe], [O/Fe] vs. [Cr/Fe], [Fe/H] vs. [$\alpha$/Fe], it shows that all of them has correlation, more or less with age, positive or negative. Especially for the last one [$\alpha$/Fe] and [Fe/H], we could see thick disk population with red patch and thin disk population with blue patch.

As we mentioned, almost all of parameters are correlated with age but why do we only choose first six to nine parameters for our method and why the other parameters shown in Fig.~\ref{Feature_importances} do not have high importance. The reason is that we find they are related to the properties of random forest method, it means that when there are correlations for multiple features, the RF will extract the one with the greatest contribution, and then the importance of other features might become not very important artificially (e.g., [Fe/H]).

\begin{figure*}
  \centering
  \includegraphics[width=0.8\textwidth]{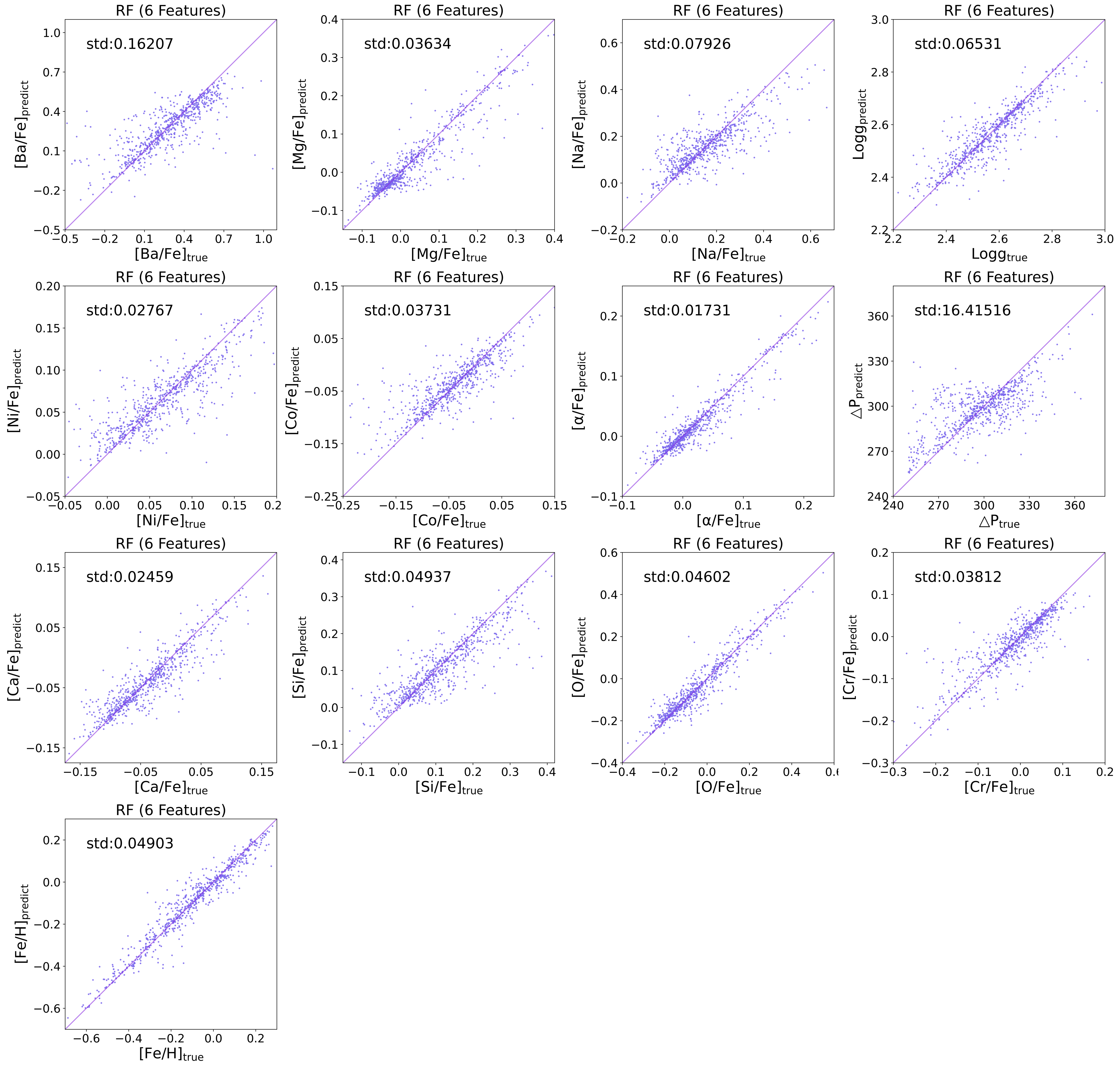}
  \caption{The predicting results for some chemical parameters using the first six stellar parameters shown in feature importance of Fig.~\ref{Feature_importances}. The consistency is quite well and  from the top left to the bottom left are: [Ba/Fe], [Mg/Fe], [Na/Fe], $logg$, [Ni/Fe], [Co/Fe], [$\alpha$/Fe], $\Delta$P, [Ca/Fe], [Si/Fe], [O/Fe], [Cr/Fe], [Fe/H].}
  \label{Feature_prediction}
\end{figure*}

As a test, we attempt to use the first six stellar parameters of importance to independently predict other stellar parameters in RC-age sample and check the predicted results. As shown in Fig.~\ref{Feature_prediction}, we find that other stellar parameters can be predicted by using the first six stellar parameters. Because the first six features are more or less related to other features, the importance of other features behave not so significant when we make the related analysis.

\begin{figure*}
  \centering
  \includegraphics[width=0.9\textwidth]{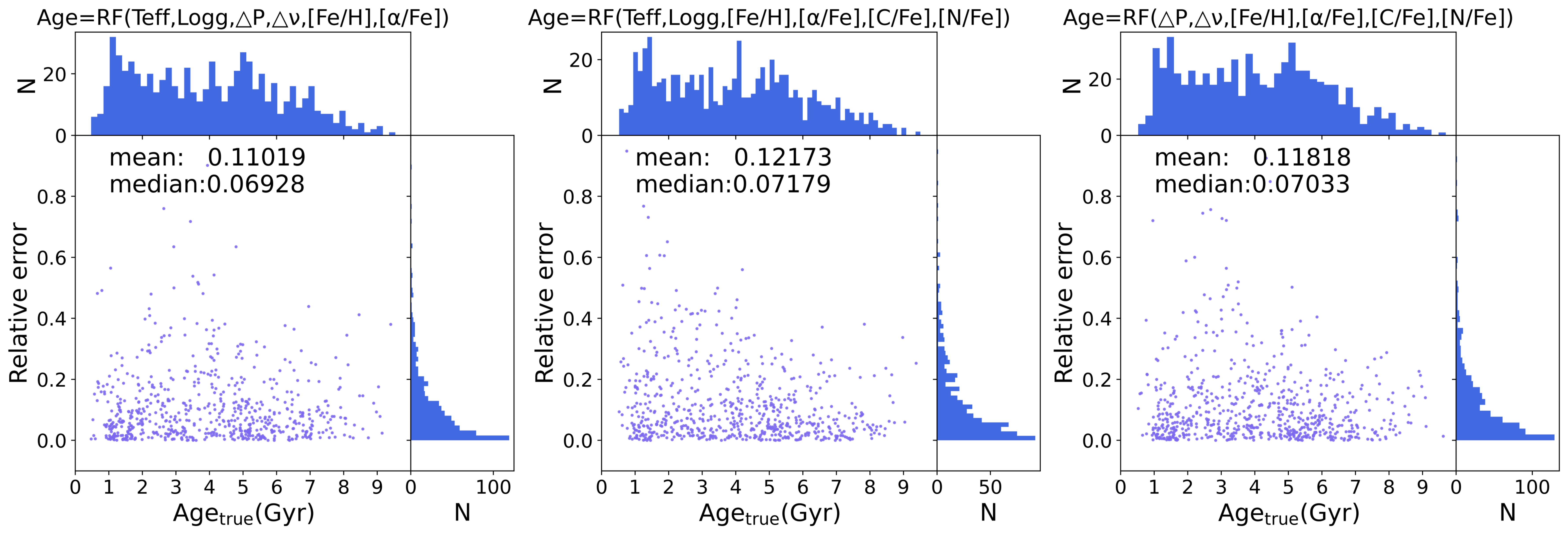}
  \caption{Age determination of RC stars for test, with six empirical stellar parameters shown in the top of each figure. We could see the precision is almost the same.}
  \label{Six_prediction}
\end{figure*}

Inversely, we also randomly choose several other relevant stellar parameters to empirically predict age in order to compare with our previous results, notice the Fig.~\ref{Six_prediction}. Obviously, we find that even though we use other parameters to predict the similar precision could be reached. All these results show that our method for age and mass estimation is reasonable and we could make full use of many parameters to estimate age and mass for other catalogs even though we are lack of some chemical stellar parameters.

\section{Discussion}

\subsection{Comparisons for age prediction using different catalogs}

\begin{figure}
  \includegraphics[width=0.45\textwidth]{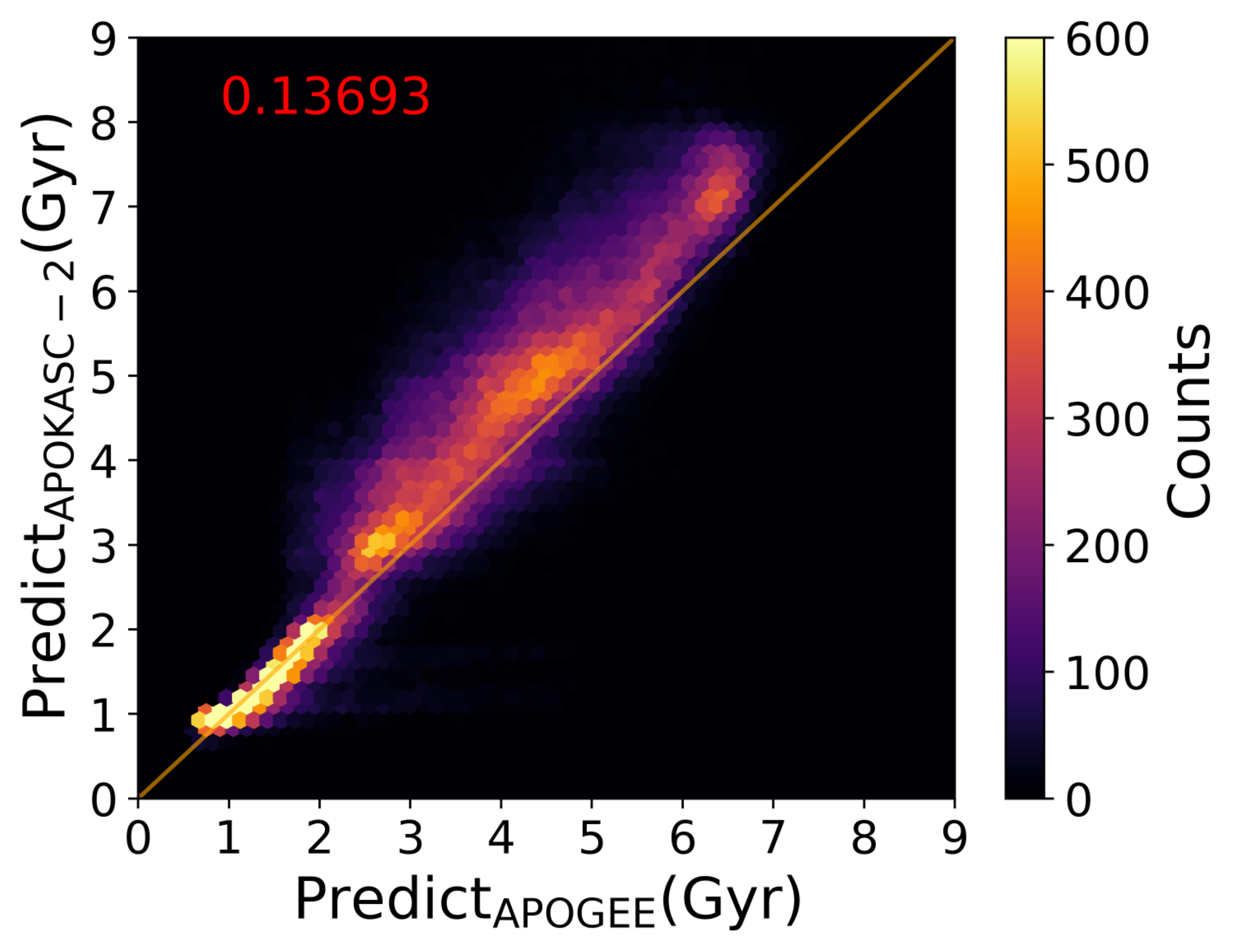}
  \caption{Comparison of age predictions using two different age catalogs. x axis is age trained by APOGEE and the y axis is trained by APOKASC-2 during our work, the figure coloured by star counts.}
  \label{APOGEE - APOKASC}
\end{figure}

\begin{figure}
  \includegraphics[width=0.65\textwidth]{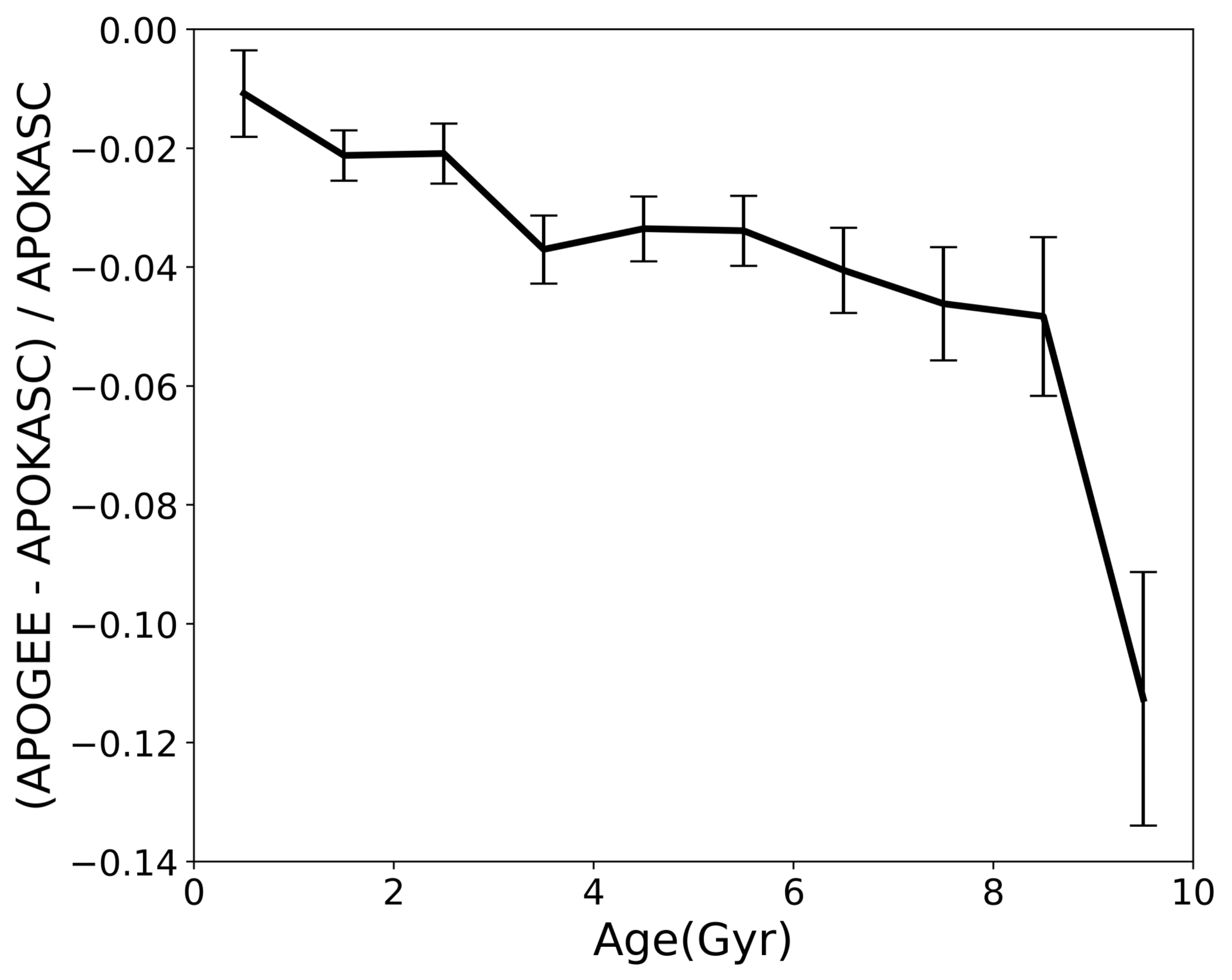}
  \caption{The relative age error of APOGEE and APOKASC-2 along with the age for common stars. We use two catalogs to make prediction and find that the older the star, the more obvious for the difference. The error bar is poission noise.}
  \label{Age - RE}
\end{figure}

In this paper, we choose the RC age of APOKASC-2 as the training dataset because it is the high resolution asteroseismology sample.  In order to compare the age based on the APOKASC-2 and APOGEE \citep{2019ApJ...879...69T}, we use these two different catalogs to predict age, as shown in Fig.~\ref{APOGEE - APOKASC}, the x-axis is age trained by APOGEE and the y-axisl is trained by APOKASC-2, they have different stellar number. We find that for older stars, the age predicted by APOKASC-2 is systematically higher than the stars predicted by APOGEE, which is caused by the different dataset precision possibly.  And as can be seen from Fig.~\ref{Age - RE}, showing the relative error analysis for these two catalogs, the age based on APOGEE is systematically smaller than that based on APOKASC-2. With the increasing of age, the difference becomes more and more obvious, however, for the overall trend, almost of all difference are within 10\% which is acceptable and implying that the precision of prediction is dependent on the quality of the dataset. 

\begin{figure}
  \centering
  \includegraphics[width=0.45\textwidth]{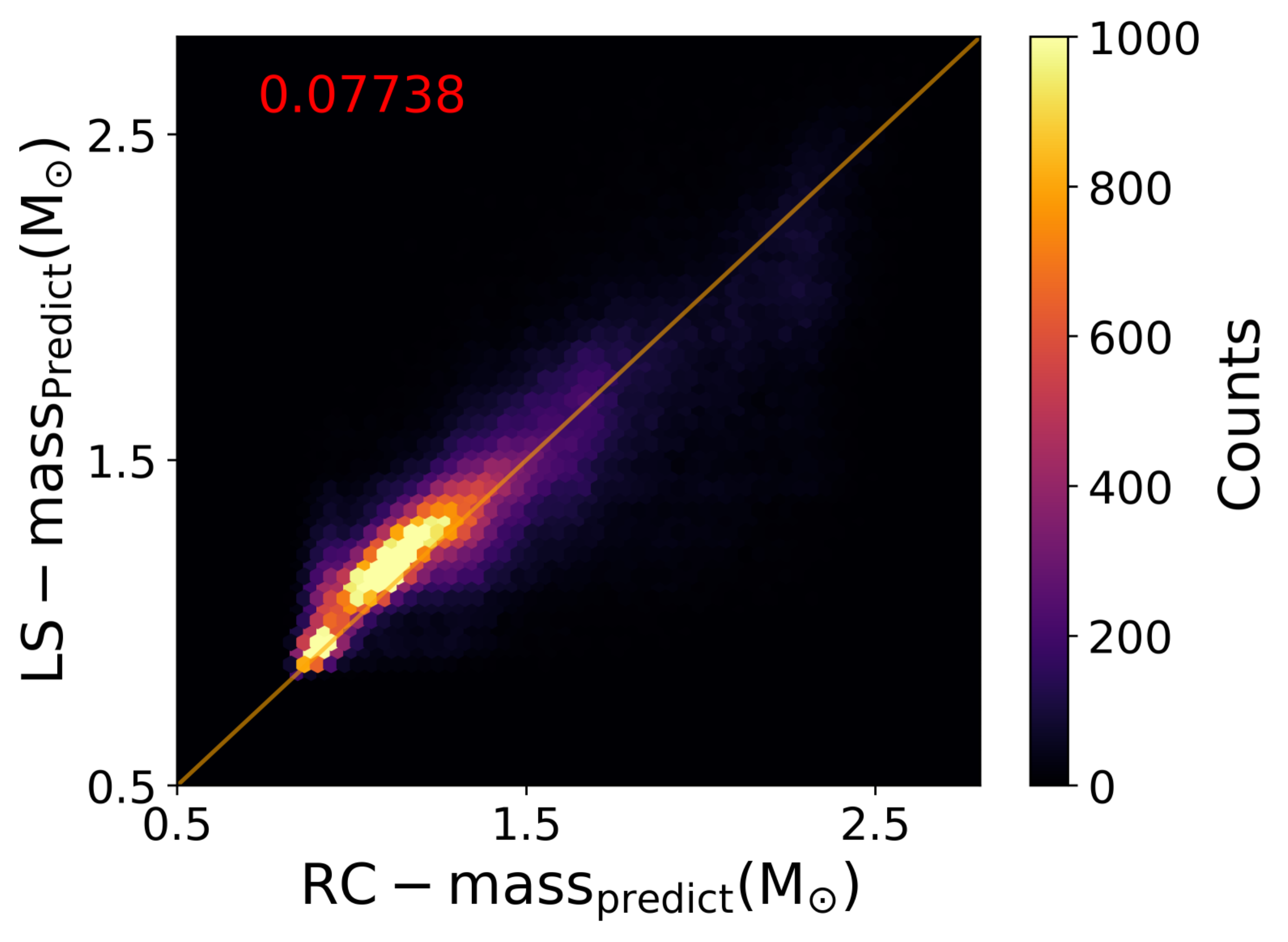}
  \caption{The comparison of common stars between the LS-mass and RC-mass in this work.}
  \label{Hexbin_LS vs RC}
\end{figure}

\subsection{Comparison of common stars between two different mass predictions during this work}

We have predicted the mass of two groups of samples, the LS-mass with convex hull algorithm and the RC-mass without convex hull algorithm. After cross match we find 155,532 common stars and then we compare the two slightly different mass prediction methods. 

As can be seen from Fig.~\ref{Hexbin_LS vs RC}, the values of relative errors is 8\%, which shows that the mass difference predicted by the two methods is small and self-consistent.

\subsection{Comparison of different machine learning methods}

Different machine learning methods used in this work have their own characteristics but there should be no absolute difference for the advantages and disadvantages, which is dependent on the specific purposes. The reason why we choose RF is that after many attempts, we find that it is better to in line with our expectations. The quantitative comparison of the six machine learning methods including bayesian linear regression (hereafter: BYS), gradient boosting decision tree (hereafter: GBDT), multilayer perceptron (hereafter: MLP), multiple linear regression (hereafter: MLR), random forest (RF) and support vector regression (hereafter: SVR) is shown in this section. 

\begin{figure*}
  \centering
  \includegraphics[width=0.9\textwidth]{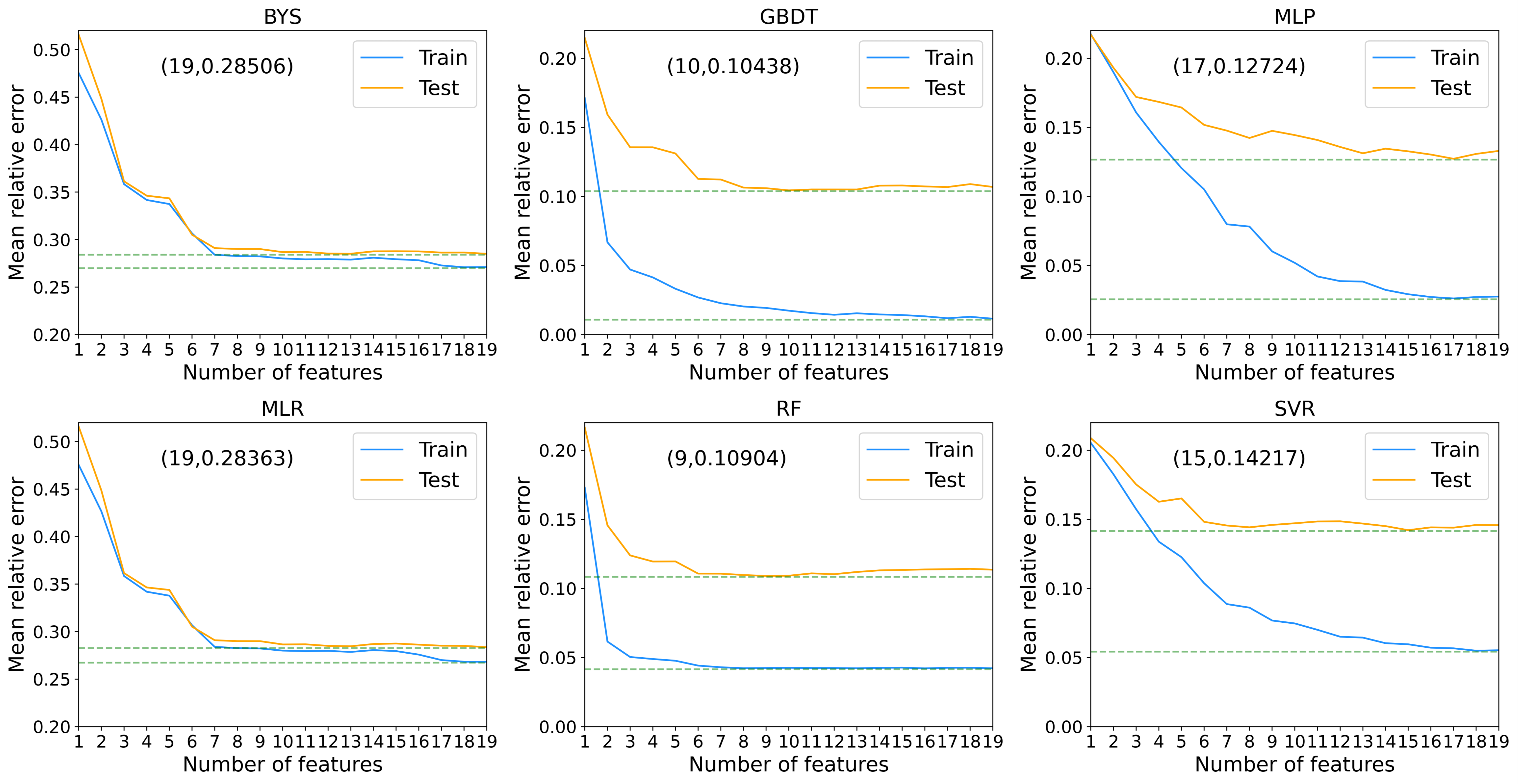}
  \caption{The relations between the number of training features and the mean relative error of the test dataset for the six prediction models. Different figures are different machine learning methods and different colour lines are training and test dataset respectively, the horizontal dashed lines are used to guide our eyes for the stable pattern. Notice that it is the minimum value labeled but it is not the final feature we adopt, we choose final features shown in Fig.~\ref{Diff_prediction} empirically and reasonably.}
  \label{Diff_NOF - MRE}
\end{figure*}

\begin{figure*}
  \centering
  \includegraphics[width=0.9\textwidth]{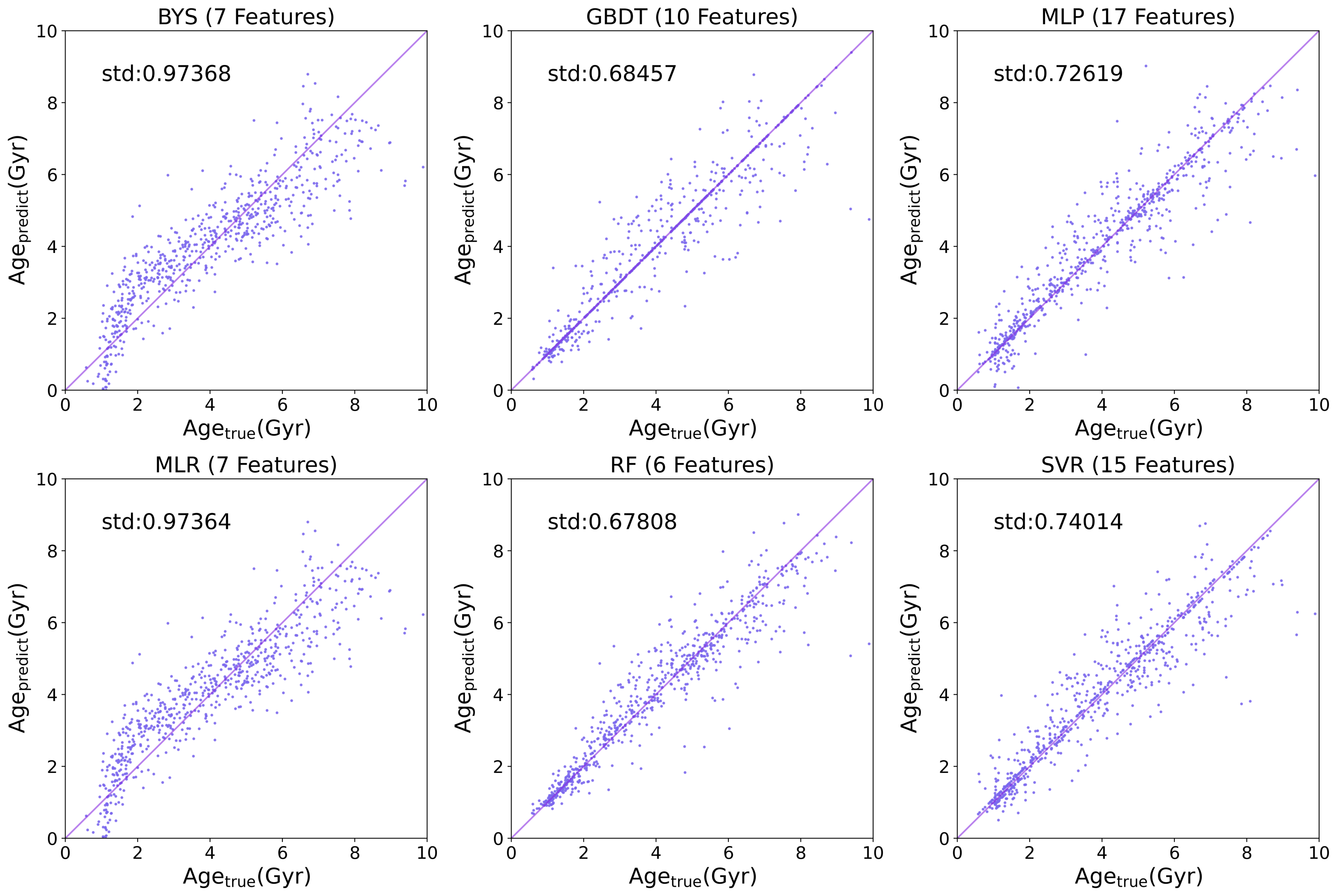}
  \caption{The comparison between our predicted ages and the true values we use in the catalog. Different figures are different machine learning methods and the dispersions are labeled in each panel and diagonal lines are used for comparison. The final number of features (corresponding to Fig.~\ref{Feature_importances}) we adopt to train each model is marked at the top of each panel.}
  \label{Diff_prediction}
\end{figure*}

Fig.~\ref{Diff_NOF - MRE} shows the relation between features number used in the training model and the median relative error in different machine learning methods. Meanwhile, Fig.~\ref{Diff_prediction} shows the age prediction of different methods for the test dataset. Based on the value labeled in panels of these two figures, we can clearly see that BYS and MLR are relatively worse because both of them have the higher median relative error of $\sim$ 28\% and larger dispersions of 0.97 \,Gyr, which might be caused by that our prediction of RC stars is nonlinear, but BYS and MLR are linear models. 

Among the other nonlinear methods, the MLP are difficult to adjust during our experiments and the performance is hard to keep stable, the median relative error is 13\% and dispersion is 0.73 \,Gyr. And the median relative error and dispersion of SVR are 14\% and 0.74 \,Gyr. We can see from Fig.~\ref{Diff_NOF - MRE}, the precision of GBDT is similar to the RF with median relative error of 10\% and dispersion of 0.68 \,Gyr, but more features of GBDT (10) are needed than the RF (6) when the median relative error is becoming stable, in order to make our trained model applicable to more stars with fewer features, we decide to choose the RF for this work. More introduction about the six machine methods will be presented in the Appendix part. 

\section{Conclusions}

In this paper, with the help of LAMOST, APOGEE and asteroseismology data, we use random forest to predict the mass of  948,216 large sample stars, mass and age of 163,105 RC stars. We select stellar parameters with high correlation with mass and age to construct training model, then we use theses features, convex hull algorithm and random forest method to determine the age and mass of larger sample. 

We find that the precision of the mass for large sample stars could reach 3\%, RC stars could reach 4\%, and RC age precision could be 7\% for test dataset (shown in Fig.~\ref{RF_prediction}). Compared with other high quality sample, the precision for mass of large sample stars could reach 13\%, mass precision of RC stars could reach 9\%, and age precision of RC stars could reach 18\% for the median relative error. In general, our results could be compared well to recent works, in particular for open clusters, which could reach 9.5\% for median relative error, so it is strongly implying we could make full use of the method in the future.

We also explore the performance of different machine learning methods for the first time, in particular for age. There should be no absolute advantages and disadvantages between different machine learning methods, and each method has its own applications dependent on purpose. After comparisons, we find that the nonlinear model is more in line with our expectations than the linear model, and the GBDT and RF are better. In order to make the model suitable for more stars, we choose the RF which needs less feature numbers to achieve our scientific target in this work.

To some extent, this paper could be considered as the first paper of our series of works and the catalog will be shared online with community. This method will be widely used in the other catalogs or surveys and we will also attempt to consider systematics, possible zero-points for age in the future.

\acknowledgements

We would like to thank the anonymous referee for his/her very helpful and insightful comments. Thanks for the helpful comments from L{\'o}pez-Corredoira Mart{\'i}n. HFW is supported by the CNRS-K.C.Wong Fellow in France and we acknowledge the science research grants from the China Manned Space Project with NO. CMS-CSST-2021-B03, CMS-CSST-2021-A08. HFW also acknowledges the support from the project “Complexity in self-gravitating systems” of the Enrico Fermi Research Center (Rome, Italy). L.Y.P is supported by the National Key Basic R \& D Program of China via 2021YFA1600401, the National Natural Science Foundation of China (NSFC) under grant 12173028, the  Chinese Space Station Telescope project: CMS-CSST-2021-A10, the Sichuan Science and Technology Program (Grant No. 2020YFSY0034), the Sichuan Youth Science and Technology Innovation Research Team (Grant No. 21CXTD0038), Major Science and Technology Project 535 of Oinghai Province (Grant No. 2019-ZJ-A10), and the Innovation Team Funds of China West Normal (Grant No. KCXTD2022-6).

H.F.W. is fighting for the plan ``Mapping the Milky Way (Disk) Population Structures and Galactoseismology (MWDPSG) with large sky surveys" in order to establish a theoretical framework in the future to unify the global picture of the disk structures and origins with a possible comprehensive distribution function. We pay our respects to elders, colleagues and others for comments and suggestions, thanks to all of them. The Guo Shou Jing Telescope (the Large Sky Area Multi-Object Firber Spectroscopic Telescope, LAMOST) is a National Major Scientific Project built by the Chinese Academy of Sciences. Funding for the project has been provided by the National Development and Reform Commission. LAMOST is operated and managed by National Astronomical Observatories, Chinese Academy of Sciences. This work has also made use of data from the European Space Agency (ESA) mission {\it Gaia} (\url{https://www.cosmos.esa.int/gaia}), processed by the {\it Gaia} Data Processing and Analysis Consortium (DPAC, \url{https://www.cosmos.esa.int/web/gaia/dpac/consortium}). Funding for the DPAC has been provided by national institutions, in particular the institutions participating in the {\it Gaia} Multilateral Agreement.

\appendix

\section{Machine learning methods introduction}

MLR is a linear model assuming that there is a simple weighted summation relations between the variable and the predicted parameter. It has good performance in some cases although the assumption is strong. BYS is applied bayesian inference to the linear regression model. The parameters in the linear model are regarded as random variables and then we could calculate the posterior distribution, it has the basic properties of bayesian statistical model. In this experiment the data will be repeated and the overfitting will be prevented effectively, but it is computationally expensive.To be more specific for some details, in this work, we don't change most of the default parameters in sklearn software for MLR and the parameter fit$_{intercept}$ is set to be true, which means we could calculate the intercept value for this model. For BYS, we set the parameter n$_{iter}$=30 and tol=1.e-3, the meaning of “n$_{iter}$” is the maximum number of iterations and “tol” setting could stop the algorithm if it has been converged.

Both RF and GBDT are based on decision trees, the difference is that the former one uses bagging and the latter one uses boosting. The final predicting result is dependent on the decision trees and is random due to the random sampling at the beginning. The correlation between different parameters can be easily identified with the help of information gain when we are training the model. Moreover, both of these methods have good robustness and overfitting could be avoided. Because the good robustness sometimes standardization or normalization might not be needed. Good robustness means that the machine learning model could have good precision for parameters. For RF, we set the parameter n$_{estimators}$=2000, n$_{jobs}$=-1, max$_{features}$='auto' and min$_{samples-leaf}$=1. Here “n$_{estimators}$” means the number of trees in the forest, “n$_{jobs}$” can change the number of jobs in order to run in parallel, “max$_{features}$” defines the maximum number of features of each tree, and “min$_{samples-leaf}$” is the minimum number of samples required at the node. For GBDT, we set the parameter n$_{estimators}$=2000, learning$_{rate}$=0.1, subsample=1.0, loss='ls', max$_{features}$ = None and min$_{samples_leaf}$=1. The “n$_{estimators}$” means the number of boosting parameter, learning$_{rate}$ is the weight contribution of each tree, “subsample” defines the stellar fraction used for fitting the individual learners, “loss” setting could optimise the loss function, and 'ls' means least squares regression in the method. 

MLP is one of neural network models consist of input layer, hidden layer and output layer respectively. Each layer is closely connected to the neurons. It is sensitive to the overfitting and difficult to adjust the parameters, the computer time is proportional to the networks. SVR's regression is dependent on the hyperplane constructed from the datasets. Since supervised learning method is based on the symmetric loss function for training, one of the advantages is that the computational complexity does not depend on the dimension of the data, but when the dimensions are more than the number of data points the results might not be acceptable. For MLP, we set the parameter hidden layer sizes = 147, it means there is only one hidden layer with 147 neurons due to that we found too complex networks would not improve the prediction performance. For SVR, the gaussian kernel has been used, and we set the parameter C=15 (regularization parameter). Notice if “C” is too large or too small, the prediction performance will be reduced. More details could be found in Scikit-learn (sklearn) publicly available package \citep{2019arXiv191006853A,2019arXiv191101217M,2020arXiv200511251F}.

\end{document}